\documentclass{article}

\usepackage[preprint]{neurips_2023}

\usepackage[utf8]{inputenc} % allow utf-8 input
\usepackage[T1]{fontenc}    % use 8-bit T1 fonts
\usepackage{hyperref}       % hyperlinks
\usepackage{url}            % simple URL typesetting
\usepackage{booktabs}       % professional-quality tables
\usepackage{amsfonts}       % blackboard math symbols
\usepackage{nicefrac}       % compact symbols for 1/2, etc.
\usepackage{microtype}      % microtypography
\usepackage{xcolor}         % colors

\usepackage{amsmath}
\usepackage{amssymb}
\usepackage{mathtools}
\usepackage{amsthm}

\usepackage{multirow}
\usepackage{bm}
\usepackage{array}

\usepackage{natbib}

\newcommand{\sJM}{\mathop{\sum\nolimits\sp{\ne}}}

\title{XGBoostPP: Tree-based Estimation of Point Process Intensity Functions}

\author{%
  Changqing~Lu$^1$ \qquad
  Yongtao~Guan$^2$\thanks{Corresponding Author. \texttt{guanyongtao@cuhk.edu.cn}} \qquad
  Marie-Colette~van~Lieshout$^{31}$ \qquad
  Ganggang~Xu$^4$ \\
  $^1$University of Twente, Enschede, The Netherlands\\
  $^2$The Chinese University of Hong Kong, Shenzhen, China\\
  $^3$Centrum Wiskunde \& Informatica, Amsterdam, The Netherlands\\
  $^4$Univeristy of Miami, Coral Gables, FL, USA
}

\begin{document}

\maketitle

\begin{abstract}
  We propose a novel tree-based ensemble method, named XGBoostPP, to nonparametrically estimate the intensity of a point process as a function of covariates. It extends the use of gradient-boosted regression trees~\citep{Chen2016XGBoost} to the point process literature via two carefully designed loss functions. The first loss is based on the Poisson likelihood, working for general point processes. The second loss is based on the weighted Poisson likelihood, where spatially dependent weights are introduced to further improve the estimation efficiency for clustered processes. An efficient greedy search algorithm is developed for model estimation, and the effectiveness of the proposed method is demonstrated through extensive simulation studies and two real data analyses. In particular, we report that XGBoostPP achieves superior performance to existing approaches when the dimension of the covariate space is high, revealing the advantages of tree-based ensemble methods in estimating complex intensity functions.
\end{abstract}

\section{Introduction}
\label{sec:introduction}

Point process models are increasingly popular for analyzing point pattern data across diverse disciplines, including tourism (e.g.,~\citealp{D’Angelo2022}), sociology (e.g.,~\citealp{xu2017,Zhang2023}), criminology (e.g.,~\citealp{yin2021fused,zhu2022}) and epidemiology (e.g.,~\citealp{dong2023,Schoenberg2023}). A crucial aspect of point process analysis is the estimation of the intensity, which defines the likelihood of event occurrences over the observation window. Existing literature explores two primary scenarios for point process intensity estimation: one based on the Cartesian coordinates of events (e.g., \citealp{Diggle1985KDEpoints}) and the other involving spatial covariates (e.g., \citealp{Guan2008KIE,Baddeley2012KIE}). This paper concentrates on the latter, emphasizing the significance of spatial covariates in explaining intensity variation. % among nearby locations. influenced by differences in their covariates, whereas distant locations may exhibit similar intensities due to shared covariates. 

Relating spatial covariates to the intensity of a point process has a well-established history. A particularly useful approach is to assume a parametric form for the intensity function, e.g. a log-linear model that formulates the log-intensity as a linear combination of accessible covariates. Extensive studies have been conducted on the estimation accuracy and asymptotic properties of the resulting estimators (see, e.g.,~\citealp{Schoenberg2005PoissonLikelihood, waagepetersen2009two, Guan2015Quasilikelihood,hessellund2022semiparametric,xu2023semiparametric}). Although a parametric model is easy to estimate and interpret, it may not be flexible enough for certain applications due to the inherent rigidity of the parametric form.

In contrast, nonparametric approaches impose fewer and less unrealistic assumptions and have gained considerable attention recently. Examples include kernel intensity estimators (KIE) \citep{Guan2008KIE, Baddeley2012KIE}, Gaussian Cox process approaches (GCP) \citep{Cunningham2008, Kim2022AugmentedPP} and Bayesian nonparametric models~\citep{Yin2022NBAstudy}. KIEs apply the standard kernel smoother in the covariate space and are computationally efficient \citep{Baddeley2015Spatstat}. However, due to high local variability, the estimation accuracy of KIEs may not be as good as other nonparametric estimators. GCPs model the latent intensity as a function of a Gaussian process and require more computational resources to approximate the intractable integration of the Gaussian process, either through MCMC algorithms \citep{Moller2004Book} or domain discretization \citep{Rue2009INLA,Illian2012}. Recent studies have proposed various approximation schemes to reduce the computational demand associated with this integration (e.g.,\ \citealp{Walder2017PermanentalPP, Aglietti2019SigmoidalGCP}), although the computational cost remains a significant challenge.

Another limitation of existing nonparametric approaches is the constraint to consider only a small number of covariates, primarily due to the well-known `curse of dimensionality' phenomenon. For instance, KIEs typically accommodate only one or two spatial covariates without imposing additional, potentially unrealistic restrictions such as isotropy. GCPs, on the other hand, typically treat the intensity as a function of spatial coordinates, thereby lacking the capability to incorporate additional covariates. A noteworthy exception is the augmented permanental process (APP) proposed in \citet{Kim2022AugmentedPP}, which extends the work of \citet{Flaxman2017poisson}. In APP, the square root-intensity is assumed to be a Gaussian process defined on the multi-dimensional covariate domain. Although, theoretically, there is no limit on the dimension of the covariate space, our numerical results suggest a significant deterioration in the performance of APP as the covariate dimension increases.

With the rapid development of data-collection technologies, an expanding array of covariates has become accessible for point pattern analysis (see, e.g.,~\citealp{Lu2023FirePrediction}). This surge underscores the necessity for an approach adept at handling the dimensionality of covariates in nonparametric intensity estimation. Tree-based models have proven to excel in predicting complex relationships between numerous covariates and the response variable (see, e.g.,~\citealp{grinsztajn2022tree}). Moreover, ensemble methods, such as random forests \citep{breiman2001random} and gradient boosting machines \citep{natekin2013gradient}, leverage the combination of multiple weak learners to build a robust learner, behaving highly effective across various tasks. Surprisingly, to the best of our knowledge, no existing work in the point process literature has employed tree-based models for intensity estimation.

To fill the gap, in this study, we introduce a novel tree-based approach, XGBoostPP, tailored for nonparametrically estimating the intensity of a point process over a high-dimensional covariate space. Numerical experiments show that our approach demonstrates comparable performance to the existing methods when only a small number of covariates are available. However, as the covariate number increases, XGBoostPP significantly outperforms other approaches that may either encounter prohibitive computational challenges or produce significantly less accurate estimates. Finally, we apply XGBoostPP to forestry and fire data, further emphasizing its effectiveness in real-world scenarios.

\section{Background and Literature}

In this section, we provide a brief overview of point process theory, along with a summary of nonparametric intensity estimation approaches in the literature.

\subsection{Point Process Theory}
\label{subsec:PPmodels}

Consider a point process $X$ defined on a bounded domain $\mathcal{S} \subset \mathbb{R}^d$. For any Borel set $\mathcal{B}\subset \mathcal{S}$, let $|\mathcal{B}|$ and $N(\mathcal{B})$ denote the area of $\mathcal{B}$ and the number of events falling in $\mathcal{B}$. Let $\mathrm{d}\bm{s}$ denote an infinitesimal ball centred at $\bm{s}\in \mathcal{S}$. Following \citet{Diggle2003Book}, we define the intensity function of $X$ as 
$
    \lambda(\bm{s})=\lim_{|\mathrm{d}s|\to 0}\mathbb{E}[N(\mathrm{d}\bm{s})]/|\mathrm{d}s|
$
and the second-order intensity function as 
\begin{equation*}
\lambda^{(2)}(\bm{x},\bm{y})=\lim_{|\mathrm{d}\bm{x}|,|\mathrm{d}\bm{y}|\to 0} \frac{\mathbb{E}\left[N(\mathrm{d}\bm{x})N(\mathrm{d}\bm{y})\right]}{|\mathrm{d}\bm{x}||\mathrm{d}\bm{y}|}.
\end{equation*}
By scaling, one obtains the pair correlation function $g(\bm{x},\bm{y}) = \lambda^{(2)}(\bm{x},\bm{y})/[\lambda(\bm{x})\lambda(\bm{y})]$. Assuming second-order intensity-reweighted stationarity \citep{Baddeley2000SecondOrderStationary}, it can be simplified to $g(r)$ at distance $r=\|\bm{x}-\bm{y}\|$. 

\paragraph{Poisson process.} The Poisson process is defined as the following: (i) for any bounded set $\mathcal{B}$, the number of events falling in it is Poisson distributed with mean $\int_\mathcal{B}\lambda(\bm{s})\mathrm{d}\bm{s}$; (ii) for disjoint bounded sets, the numbers of events falling in them are independent. Under these assumptions, the likelihood of a realization $\bm{\mathrm{x}}$ from a Poisson process is of the form
\begin{equation*}
    \prod_{\bm{x}\in \bm{\mathrm{x}}}\lambda(\bm{x})\exp\left[-\int_\mathcal{S}\lambda(\bm{s})\mathrm{d}\bm{s}\right].
\end{equation*}
We remark that, for a Poisson process, $g(r)\equiv 1$. 

\paragraph{Log-Gaussian Cox Process.} A Cox process is a Poisson process with a random intensity function. In particular, we consider the log-Gaussian Cox process whose log-intensity is of the form
$    \log[\Lambda(\bm{s})]=\log[\Tilde{\lambda}(\bm{s})]+Y(\bm{s})$,
where $\Tilde{\lambda}(\cdot)$ is a deterministic function of spatial covariates and $Y(\cdot)$ is a zero-mean Gaussian process with a covariance kernel $\rho(\cdot,\cdot)$. For instance, $\rho(\cdot,\cdot)$ can take, but is not limited to, the exponential form
$\rho(\bm{x},\bm{y})=\tau^2\exp\left(-\|\bm{x}-\bm{y}\|/\sigma\right)$
with $\tau^2$ as the variance and $\sigma$ as the scaling parameter. The intensity of such a process is given by $\lambda(\bm{s})=\mathbb{E}[\Lambda(\bm{s})]=\Tilde{\lambda}(\bm{s})\exp(\tau^2/2)$, and the pair correlation function reads
$g(r)=\exp\left[\tau^2\exp\left(-r/\sigma\right)\right]$ which quickly approaches one as $r$ becomes large.

\paragraph{Neyman-Scott Process.} A Neyman-Scott process is generated in two stages. First, a set of parents, denoted by $\bm{\mathrm{c}}$, are produced from a homogeneous Poisson process of intensity $\kappa$. Then, for each parent $\bm{a}\in\bm{\mathrm{c}}$, a set of offspring points, denoted by $X_{\bm{a}}$, are generated independently from a Poisson process with an intensity function
$
    \lambda_{\bm{a}}(\bm{s})=k(\bm{s}-\bm{a};\sigma)\Tilde{\lambda}(\bm{s}),
$
where $k(\cdot;\sigma)$ is a density function parameterized by a spreading scalar $\sigma$. In particular, the Thomas process assumes $k(\cdot;\sigma)$ to be an isotropic Gaussian kernel with a standard deviation $\sigma$. As a result, the overall intensity function of such a process is given by $\lambda(\bm{s})=\kappa\Tilde{\lambda}(\bm{s})$, and the pair correlation function reads
$
    g(r)=1+\exp(-r^2/4\sigma^2)/(4\pi\kappa\sigma^2)
$
which also quickly approaches one as $r$ grows large.

\subsection{Nonparametric Estimation Approaches}
\label{subsec:approaches}

\paragraph{Kernel Intensity Estimators.} Kernel intensity estimators \citep{Guan2008KIE,Baddeley2012KIE} are extensions of kernel density estimators (KDE) \citep{Silverman2018Book} to point pattern data. Similar to KDEs, estimation can be challenging for KIEs when the dimension of the covariate space is high (see, e.g.,\ \citealp{Ottmar2018Bandwidth}). To address this challenge, several dimension-reduction tools have been developed (e.g.,~\citealp{Guan2010KIEDimension}). Implementations of KIEs are available in the R-package \textit{spatstat} \citep{Baddeley2015Spatstat}, where only data with at most two covariates are supported.

\paragraph{Gaussian Cox Process Approaches.} Gaussian Cox process approaches \citep{Cunningham2008,Adams2009SigmoidalGCP,Gunter2014,Samo2014,Lloyd2015PermanentalPP,Walder2017PermanentalPP,Donner2018SigmoidLGCP,John2018LargeScaleGCP,Aglietti2019SigmoidalGCP,Kim2022AugmentedPP} can be viewed as Bayesian alternatives to KIEs. GCPs model the intensity by a Gaussian process with a positive link function and maximize the posterior probability. Different link functions lead to variants of GCPs, such as log-Gaussian Cox processes (exponential link; \citealp{Moller1998LGCP}), permanental processes (quadratic link; \citealp{McCullagh2006PermanentalP}) and sigmoidal Gaussian Cox processes (sigmoid link; \citealp{Adams2009SigmoidalGCP}). To our knowledge, most GCPs estimate the intensity as a function of spatial coordinates, with the exception of the augmented permanental process approach \citep{Kim2022AugmentedPP}. APP assumes that the square root-intensity is generated from a Gaussian process defined on the covariate domain and tackles the intractable integration via path integral formulation, achieving great computational improvements. However, it is not immune to the curse of dimensionality. In our numerical experiments, a significant deterioration in its performance is observed as the dimension of the covariate space increases.

\section{Methodology}

\subsection{The XGBoostPP Model}
\label{subsec:XGBoostPP}

Consider a point process $X$ defined on an observation window $\mathcal{S}$. Assume that its intensity $\lambda$ is a function of a $p$-dimensional spatial covariate vector $\bm{z}(\bm{s})$ with $\bm{s}\in\mathcal{S}$. To estimate the complex mapping $\lambda[\bm{z}(\bm{s})]:\mathbb{R}^p \to \mathbb{R}^+$, we propose a tree-based ensemble model, XGBoostPP, following the classic scalable tree boosting system -- XGBoost~\citep{Chen2016XGBoost}. The standard XGBoost was designed for general regression or classification tasks and is not directly applicable to point pattern data with specific spatial structures. Hence, we propose two carefully designed likelihood-based loss functions and an efficient learning algorithm to adapt to the point process context.

Formally, our XGBoostPP model estimates the log-intensity of $X$ using $K$ additive trees:
\begin{equation}
    \log\left\{\lambda\left[\bm{z}(\bm{s})\right]\right\}=\sum_{k=1}^K f_k\left[\bm{z}(\bm{s})\right], \quad f_k\in \mathcal{F},
    \label{e:1}
\end{equation}
where $\mathcal{F}=\{f[\bm{z}(\bm{s})]=\boldsymbol{\theta}_{q[\bm{z}(\bm{s})]}\}$ is the space of regression trees with $q[\bm{z}(\bm{s})]: \mathbb{R}^p \to \{1,\dots, N_q\}$ and $\boldsymbol{\theta}_{q[\bm{z}(\bm{s})]}\in \mathbb{R}^{N_q}$. Here, $q$ represents a tree structure working as a function that maps the covariate information at a location $\bm s$ to a corresponding leaf index, $N_q$ is the number of leaves in this tree and $\boldsymbol{\theta}_{q}$ is the vector of leaf scores. Each tree $f_k$ is uniquely defined by an independent tree structure $q_k$ and the associated leaf score vector $\boldsymbol{\theta}_{q_{k}}$. In the remainder, we denote the estimated intensity function as $\lambda(\bm{x};\bm{f})$ with $\bm f=(f_1,\cdots,f_K)$ and write $\boldsymbol{\theta}_k$ for $\boldsymbol{\theta}_{q_{k}}$ with $\theta_{kv}$ representing the element of $\boldsymbol{\theta}_k$ on leaf $v$. Given a location $\bm{s}$, the decision rules of the trees in $\bm{f}$ first classify $\bm{z}(\bm{s})$ into $K$ leaves and then calculate the estimated log-intensity by adding up the scores on all corresponding leaves.

%An XGBoostPP with given tree structures $\{q_i\}_{i=1}^K$ can be interpreted as a generalized linear model with `transformed' covariates. Specifically, the original covariates $\bm{z}(\bm{s})$ are replaced by the membership identities of a point $\bm{s}$ with respect to the leaves of all trees. If a point belongs to a leaf in a tree, the corresponding element in the membership vector is assigned one; zero otherwise. Denote the new covariate vector by $\bm{Z}(\bm{s})$ and the concatenated score vector of all trees by $\boldsymbol{\theta}=[\boldsymbol{\theta}_1,\dots,\boldsymbol{\theta}_K]$. The intensity function can be formulated as 
%$
%    \lambda[\bm{Z}(\bm{s});\boldsymbol{\theta}]=\exp[\boldsymbol{\theta}^{\top}\bm{Z}(\bm{s})].
%$
%Such reparameterization allows XGBoostPP to exploit the utility of the weighted Poisson likelihood loss, which will be discussed in Section~\ref{subsec:WPLloss}. For simplicity, below, we write $\lambda(\bm{s})$ for $\lambda[\bm{Z}(\bm{s})]$ and $\lambda(\bm{s};\boldsymbol{\theta})$ for $\lambda[\bm{Z}(\bm{s});\boldsymbol{\theta}]$.

\subsection{Likelihood-based Loss Functions}
\label{subsec:LossFunctions}

To estimate the structures of the $K$ tree predictors in the XGBoostPP model, as well as their leaf scores, we propose two likelihood-based loss functions: the Poisson likelihood loss and the weighted Poisson likelihood loss.

\subsubsection{Poisson Likelihood Loss} 

It is well-known that a parametric intensity function defined as $\lambda(\bm{s};\boldsymbol{\beta})=\exp [\boldsymbol{\beta}^{\top}\bm{z}(\bm{s})]$ can be consistently estimated by maximizing the Poisson log-likelihood~\citep{Schoenberg2005PoissonLikelihood}
\begin{equation*}
\sum_{\bm{x}\in \bm{\mathrm{x}}}\log\left[\lambda(\bm{x};\bm\beta)\right]-\int_{\mathcal{S}}\lambda(\bm{s};\bm\beta)\mathrm{d}\bm{s},
\end{equation*}
even when the underlying point process is not Poisson. Motivated by this phenomenon, we propose to approximate the optimal tree predictors $\bm{f}$ in model \eqref{e:1} by minimizing the following penalized loss function
\begin{equation}
\begin{split}
    L(\bm f)=\sum_{k=1}^K\Omega(f_k)-\sum_{\bm{x}\in \bm{\mathrm{x}}}\phi(\bm{x};\bm f)
    +\int_{\mathcal{S}}\exp\left[\phi(\bm{s};\bm f)\right]\mathrm{d}\bm{s},
    \label{e:2}
\end{split}
\end{equation}
where $\phi(\bm{s};\bm f)=\sum_{k=1}^K f_k\left[\bm{z}(\bm{s})\right]$ and $\Omega(f_k)=\gamma\sum_v |\theta_{kv}|$ with some $\gamma>0$. The last two terms of~\eqref{e:2} correspond to the negative Poisson log-likelihood and the first term penalizes the complexity of $\bm{f}$ to avoid overfitting.

Note that one needs to minimize~\eqref{e:2} with respect to both tree structures, i.e. $\{q_k\}_{k=1}^K$, and leaf scores, i.e. $\{\bm\theta_k\}_{k=1}^K$, simultaneously. However, it is impractical to exhaust all possible tree structures. Thus, for feasible computation, we develop a greedy additive search algorithm, as illustrated in Sec.~\ref{subsec:TrainingManner}, for the optimal intensity estimate.

%Using the language of Fan and Li (2001), we assume an oracle estimator of the intensity function under the XGBoostPP model exists, which knows the true supporting leaves, denoted by $\bm{Z}^{*}(\bm{s})$. Minimizing ~\eqref{e:2} then yields a LASSO problem, which can be regarded as approximating the oracle estimator from a large transformed covariate space including all possibilities of $\bm{Z}(\bm{s})$ by pruning the useless leaves (setting the scores $\theta$ on them to 0) and estimating the optimal $\theta$ for the true supporting leaves. 

%Considering the oracle tree structures $\bm{Z}^{*}(\bm{s})$, minimizing ~\eqref{e:2} for an underlying Poisson process model returns unbiased and asymptotically efficient estimates of $\theta$ if $\gamma$ is properly selected and the asymptotic regime is appropriately defined (see, \cite{Choiruddin2018}).

%Suppose that we have achieved an XGBoostPP of certain tree structures. The efficiency of the estimates of $\boldsymbol{\theta}$ for a Poisson $X$ is given by the proposition below.
%\begin{proposition}
%    If $X$ is a Poisson process, conditional on $\bm{Z}(\bm{s})$, the estimates of corresponding leaf scores $\boldsymbol{\theta}$ obtained by minimizing Eq.~(\ref{e:2}) are asymptotically most efficient.
%\end{proposition}
%The proof is provided in Appendix~\ref{sec:Appendix_A}. 

\subsubsection{Weighted Poisson Likelihood Loss} 

When the underlying point process deviates from a Poisson process, e.g. the log-Gaussian Cox process and the Neyman-Scott process, maximizing the Poisson log-likelihood, as discussed in~\citet{Schoenberg2005PoissonLikelihood}, still results in a consistent estimator for a parametric intensity function. However, the estimation efficiency can be poor due to spatial dependence \citep{Guan2010WeightedPoissonLoss}. Such a limitation also applies to the estimation of our XGBoostPP model when minimizing \eqref{e:2}. To improve the estimation efficiency, we propose a weighted Poisson likelihood loss, inspired by the quasi-likelihood developed in \citet{Guan2015Quasilikelihood}.

To motivate our proposal, %using the language of \citet{Fan2001Lasso}, 
we assume that the optimal tree structures under model \eqref{e:1} exist, denoted by $\bm{q}^{\mathcal{O}}$, and examine the `oracle' estimator for the associated optimal leaf scores that minimizes $\mathbb{E}[L(\bm f)]$ in \eqref{e:2}. In this case, all values of $\theta_{kv}$ that do not correspond to a `terminal' node in $\bm{q}^{\mathcal{O}}$ are precisely set to zero. Our focus then shifts to the estimation efficiency of the scores on the remaining nodes, denoted by $\bm\theta_{\mathcal{A}}$ with $\mathcal{A}$ indicating the set of the true supporting leaves.

By the Karush–Kuhn–Tucker optimality condition, the estimate of $\bm\theta_{\mathcal{A}}$ obtained by minimizing~\eqref{e:2} must satisfy the estimating equation
\begin{equation*}
    \bm{e}_{\bm{\phi}}(\boldsymbol{\theta}_{\mathcal{A}})
    =\gamma \mathrm{sgn}(\boldsymbol{\theta}_{\mathcal{A}})+\sum_{\bm{x}\in \bm{\mathrm{x}}}\frac{\partial \phi(\bm x;\boldsymbol{\theta}_{\mathcal{A}})}{\partial\boldsymbol{\theta}_{\mathcal{A}}}
    -\int_{\mathcal{S}}\frac{\partial \phi(\bm x;\boldsymbol{\theta}_{\mathcal{A}})}{\partial\boldsymbol{\theta}_{\mathcal{A}}}\lambda(\bm{s};\boldsymbol{\theta}_{\mathcal{A}})\mathrm{d}\bm{s}=\bm{0}.
\end{equation*}
Note that we change $\phi(\bm x;\bm{f})$ to $\phi(\bm x;\boldsymbol{\theta}_{\mathcal{A}})$ due to the knowledge of $\bm{q}^{\mathcal{O}}$. According to \citet{Guan2010WeightedPoissonLoss}, a direct generalization of $\bm{e}_{\bm{\phi}}(\boldsymbol{\theta}_{\mathcal{A}})=\bm 0$ is to consider the following more general estimating equation
\begin{equation}
\label{e:3}
    \bm{e}_{\bm{h}}(\boldsymbol{\theta}_{\mathcal{A}})
    =\gamma \mathrm{sgn}(\boldsymbol{\theta}_{\mathcal{A}})+\sum_{\bm{x}\in \bm{\mathrm{x}}}\bm{h}(\bm{x};\boldsymbol{\theta}_{\mathcal{A}})
    -\int_{\mathcal{S}}\bm{h}(\bm{s};\boldsymbol{\theta}_{\mathcal{A}})\lambda(\bm{s};\boldsymbol{\theta}_{\mathcal{A}})\mathrm{d}\bm{s}
    =\bm 0,
\end{equation}
where $\bm{h}(\bm{s};\boldsymbol{\theta}_{\mathcal{A}})$ can be any measurable function defined on $\mathcal{S}$ of the same dimension as $\boldsymbol{\theta}_{\mathcal{A}}$. \citet{Guan2015Quasilikelihood} show that the optimal $\bm{h}^*(\bm{s};\boldsymbol{\theta}_{\mathcal{A}})$ that minimizes the inverse Godambe information of the resulting estimator must satisfy the Fredholm integral equation of the second kind~\citep{Hackbusch1995Book}, that is, for any $\bm s\in\mathcal{S}$,
\begin{equation*}
    \int_{\mathcal{S}}\bm{h}^*(\bm{u};\boldsymbol{\theta}_{\mathcal{A}})\lambda(\bm{u};\boldsymbol{\theta}_{\mathcal{A}})
    \left[g(\bm{s}-\bm{u})-1\right]\mathrm{d}\bm{u}
    +\bm{h}^*(\bm{s};\boldsymbol{\theta}_{\mathcal{A}})=\nabla\lambda(\bm{s};\boldsymbol{\theta}_{\mathcal{A}})/\lambda(\bm{s};\boldsymbol{\theta}_{\mathcal{A}}),
\end{equation*}
where $\nabla\lambda(\bm{s};\boldsymbol{\theta}_{\mathcal{A}})=\partial \lambda(\bm{s};\boldsymbol{\theta}_{\mathcal{A}})/\partial \boldsymbol{\theta}_{\mathcal{A}}$. However, solving for $\bm{h}^*(\bm{s};\boldsymbol{\theta}_{\mathcal{A}})$ requires either numerical quadrature approximation \citep{Nystrom1930NystromMethod} or stochastic approximation \citep{Xu2019StochasticMethod}; both are computationally expensive. 

Alternatively, \citet{Guan2010WeightedPoissonLoss} consider the special case when
\begin{equation*}
    \int_{\mathcal{S}}\bm{h}^*(\bm{u};\boldsymbol{\theta}_{\mathcal{A}})\lambda(\bm{u};\boldsymbol{\theta}_{\mathcal{A}})
    \left[g(\bm{s}-\bm{u})-1\right]\mathrm{d}\bm{u}
    \simeq \bm{h}^*(\bm{s};\boldsymbol{\theta}_{\mathcal{A}})\lambda(\bm{s};\boldsymbol{\theta}_{\mathcal{A}}) \int_{\mathcal{S}}
    \left[g(\bm{s}-\bm{u})-1\right]\mathrm{d}\bm{u},
\end{equation*}
assuming that $\bm{h}^*(\bm{u};\boldsymbol{\theta}_{\mathcal{A}})\lambda(\bm{u};\boldsymbol{\theta}_{\mathcal{A}})\simeq \bm{h}^*(\bm{s};\boldsymbol{\theta}_{\mathcal{A}})\lambda(\bm{s};\boldsymbol{\theta}_{\mathcal{A}})$ for proximate pairs of locations and $g(\bm{s}-\bm{u})\simeq 1$ for distant pairs, which thus yields a solution  $\bm{h}^*(\bm{s};\boldsymbol{\theta}_{\mathcal{A}})=w^*(\bm{s};\bm\theta_{\mathcal{A}})\nabla\lambda(\bm{s};\boldsymbol{\theta}_{\mathcal{A}})/\lambda(\bm{s};\boldsymbol{\theta}_{\mathcal{A}})$ with a weight function
\begin{equation*}
    w^*(\bm{s};\bm\theta_{\mathcal{A}})=\frac{1}{1+\lambda(\bm{s};\bm\theta_{\mathcal{A}})\int_{\mathcal{S}}\left[g(\bm{s}-\bm{u})-1\right]\mathrm{d}\bm{u}}.
\end{equation*}
It is straightforward to show that solving \eqref{e:3} with $\bm{h}^*(\bm{s};\boldsymbol{\theta}_{\mathcal{A}})$ is equivalent to minimizing the penalized weighted negative Poisson log-likelihood
\begin{equation*}
\gamma \|\bm\theta_{\mathcal{A}}\|_1-\sum_{\bm{x}\in \bm{\mathrm{x}}}w^*(\bm{s};\bm\theta_{\mathcal{A}})\log\left[\lambda(\bm{x};\bm\theta_{\mathcal{A}})\right]
+\int_{\mathcal{S}}w^*(\bm{s};\bm\theta_{\mathcal{A}})\lambda(\bm{s};\bm\theta_{\mathcal{A}})\mathrm{d}\bm{s},    
\end{equation*}
where $\|\cdot\|_1$ indicates the $L^1$-norm of a vector.

The optimal tree structures $\bm{q}^{\mathcal{O}}$ are, of course, inaccessible in practice, but the derivation above provides useful insights into reducing the estimation variance of the leaf scores in general. Consequently, we can define the following penalized weighted loss function for our XGBoostPP model
\begin{equation}
    L_w(\bm f)=\sum_{k=1}^K\Omega(f_k)-\sum_{\bm{x}\in \bm{\mathrm{x}}}w(\bm{x};\bm f)\phi(\bm{x};\bm f)
    +\int_{\mathcal{S}}w(\bm{s};\bm f)\exp\left[\phi(\bm{s};\bm f)\right]\mathrm{d}\bm{s},
\label{e:4}
\end{equation}
where we set
\begin{equation}
    w(\bm{s};\bm f)=\frac{1}{1+\exp\left[\phi(\bm{s};\bm f)\right]\int_{\mathcal{S}}\left[g(\bm{s}-\bm{u})-1\right]\mathrm{d}\bm{u}}.
    \label{e:5}
\end{equation}
In practice, we may approximate $\int_{\mathcal{S}}[g(\bm{s}-\bm{u})-1]\mathrm{d}\bm{u}\simeq \int_{0}^m 2\pi r[g(r)-1]\mathrm{d}r$ for some sufficiently large distance $m$. The latter equals $K(m)-\pi m^2$ where $K(\cdot)$ is Ripley’s K-function (see, e.g.,~\citealp{Lieshout2019SpatialStatistics}).

%A few remarks are appropriate: the $L_1$-regularization scalar $\gamma$ does not factor into $w^*(\bm{s})$ and can be freely selected as introduced in Sec.~\ref{subsec:hyperparameterselection}. Technically, the weights needs to be nonnegative in implementation so that the weighted Poisson likelihood loss is temporarily inapplicable for repulsive processes. In addition, in \citet{Guan2010WeightedPoissonLoss}, the $\lambda(\bm{s})$ in Eq.~(\ref{e:3}) was set to $\lambda(\bm{s};\boldsymbol{\theta})$ due to their fixed covariate space of $\bm{z}(\bm{s})$. We have to consider the true intensity directly because XGBoostPP has dynamic $\bm{Z}(\bm{s})$. Concrete approximation of the weights will be particularly discussed in Sec.~\ref{subsec:dynamicWeights}. Lastly, APP also needs to solve a Fredholm integral equation \cite{Kim2022AugmentedPP}, however, we address it in a more efficient way. The detailed derivation of the weighted Poisson likelihood loss can be found in Appendix~\ref{sec:Appendix_B}.

\subsection{Additive Training Algorithm}
\label{subsec:TrainingManner}

Given the impracticality of exhausting all possible tree structures in the minimization of equations \eqref{e:2} and \eqref{e:4}, we adopt a sequential exploration in the space of $\bm{f}$, following the methodology of the standard XGBoost in~\citet{Chen2016XGBoost}. To maintain generality, we focus on the minimization of \eqref{e:4} in the subsequent demonstration.

\subsubsection{Adding Regression Trees} 

Denote the estimated log-intensity over $k-1$ trees $\hat{\bm f}_{k-1}=\{\hat f_1,\cdots,\hat f_{k-1}\}$ by $\hat{\phi}(\bm{s};\hat{\bm f}_{k-1})$ with $\bm{s}\in \mathcal{S}$. At the $k$-th iteration, we add a tree predictor $\hat f_k$ to minimize
\begin{equation*}
    L_w^{(k)}(f_k)
    =\Omega(f_k)-\sum_{\bm{x}\in \bm{\mathrm{x}}}w_{k}(\bm{x})\left[\hat{\phi}(\bm{x};\hat{\bm f}_{k-1})+f_k(\bm{x})\right]
    +\int_{\mathcal{S}}w_{k}(\bm{s})\exp\left[\hat{\phi}(\bm{s};\hat{\bm f}_{k-1})+f_k(\bm{s})\right]\mathrm{d}\bm{s},
\end{equation*}
where we iteratively update the spatial weight $w_{k}(\bm{s})$ to better approximate \eqref{e:5} according to
\begin{equation}
    \hat{w}_k(\bm{s})=\frac{\omega_k}{1+\exp\left[\hat{\phi}(\bm{s};\hat{\bm f}^{k-1})\right]\left[\hat{K}(m)-\pi m^2\right]}.
    \label{e:6}
\end{equation}
%Essentially, we greedily add a residual $f_k$ that reduces the penalized weighted Poisson likelihood loss most. 
In practice, the normalizing scalar $\omega_k$ is chosen such that  $\int_{\mathcal{S}} \hat{w}_k(\bm{s})\mathrm{d}\bm{s}=1$ and Ripley's K-function is estimated by a standard nonparametric estimator (see, e.g.,~\citealp{Moller2004Book})
\begin{equation*}
    \hat{K}(m)=\sJM_{\bm{x}_1,\bm{x}_2\in \bm{\mathrm{x}}}\frac{1\{\left\|\bm{x}_1-\bm{x}_2\|_2\leq m\right\}}{\hat{\lambda}(\bm{x}_1;\hat{\bm f}^{\mathcal{P}})\hat{\lambda}(\bm{x}_2;\hat{\bm f}^{\mathcal{P}})\left|\mathcal{S}\cap \mathcal{S}_{\bm{x}_1-\bm{x}_2}\right|},
\end{equation*}
where $\bm{x}_1, \bm{x}_2$ denote pairs of distinct points, $\|\cdot\|_2$ is the $L_2$ norm of a vector, $\hat{\lambda}(\cdot;\hat{\bm f}^{\mathcal{P}})$ represents the estimated intensity by XGBoostPP fitted using the Poisson likelihood loss \eqref{e:2} and $1/|\mathcal{S}\cap \mathcal{S}_{\bm{x}_1-\bm{x}_2}|$ is the translation edge correction factor \citep{Ohser1981EdgeCorrection}. Note that, we manually set $\hat{K}(m)-\pi m^2$ to zero when it is negative. 

By quickly applying a quadratic approximation to $L_w^{(k)}(f_k)$ and removing the constant terms involving $\hat{\phi}(\bm{s};\hat{\bm f}_{k-1})$, the loss function at the $k$-th iteration can be simplified to
\begin{equation}
    \Tilde{L}_w^{(k)}(f_k)
    =\Omega(f_k)-\sum_{\bm{x}\in \bm{\mathrm{x}}}\hat{w}_{k}(\bm{x})f_k(\bm{x})
    +\int_{\mathcal{S}}\hat{w}_{k}(\bm{s})
    \times\exp\left[\hat{\phi}(\bm{s};\hat{\bm f}_{k-1})\right]\left[f_k(\bm{s})+\frac{1}{2}f^2_k(\bm{s})\right]\mathrm{d}\bm{s}.
\label{e:7}
\end{equation}
To approximate the above integral, we use a numerical quadrature approximation; see Appendix~\ref{sec:Appendix_B} for details.

\subsubsection{Growing an Added Tree} 

To grow the newly added tree $f_k$, we follow the classic top-down greedy search algorithm. Starting from the root node, we iteratively identify the optimal split among various candidate splits until a predefined convergence criterion is reached. The key to developing a computationally feasible algorithm is thus to efficiently compute the loss reduction upon any node split.

To achieve this goal, we show in Appendix~\ref{sec:Appendix_B} that, for any given tree structure $q_k$, the optimal score $\hat \theta_{kv}$ on a leaf $v$, minimizing \eqref{e:7}, has a closed-form expression
\begin{equation*}
\hat\theta_{kv} = \frac{\mathrm{sgn}(R_{kv}-T_{kv})\max\left(\left|R_{kv}-T_{kv}\right|- \gamma,0\right)}{T_{kv}},
\end{equation*}
where we use $I_{kv}=\{\bm{s}:q_k(\bm{s})=v\}$ to denote the set of locations where $\bm{z}(\bm{s})$ belongs to leaf $v$ and
\begin{equation*}
    R_{kv}=\sum_{\bm{x}\in \bm{\mathrm{x}}}\hat{w}_{k}(\bm{x})1\{\bm{x}\in I_{kv}\},\qquad T_{kv}=\int_{\mathcal{S}}\hat{w}_{k}(\bm{s})\exp\left[\hat{\phi}(\bm{s};\hat{\bm f}_{k-1})\right]1\{\bm{s}\in I_{kv}\}\mathrm{d}\bm{s}.
\end{equation*}
It is important to note that, when $\hat\theta_{kv}=0$, this leaf will be eliminated from the tree. Hence, compared to the original weighted Poisson likelihood loss, the $L_1$-penalized loss \eqref{e:4} tends to produce a smaller tree.

Consequently, for any given tree structure $q_k$, one can compute a performance score by plugging in all optimal leaf scores as above into the loss function \eqref{e:7}. Utilizing these performance scores, one can find the best split that minimizes \eqref{e:7} among various candidate splits for each node and grow the tree sequentially. To exhaust the possible candidate splits at a node, we use the exact greedy algorithm illustrated in Algorithm~1 of \citet{Chen2016XGBoost}.

\subsubsection{More Stable Learning} 

To achieve a more stable learning performance, we also incorporate the shrinkage and subsampling techniques of the standard XGBoost~\citep{Chen2016XGBoost}. The former shrinks the impacts of the newly added tree by a parameter $\eta>0$, which is known as the learning rate in stochastic optimization problems. By replacing $f_k$ with $\eta f_k$, one leaves more room for future trees to improve model performance. The latter randomly subsamples spatial covariates as the candidate dependent variables for each node split, leading to a boosted random forest.

\subsection{Hyperparameter Selection}
\label{subsec:hyperparameterselection}
Selecting proper hyperparameters, including the number of tree predictors $K$, the learning rate $\eta$ and the penalty scalar $\gamma$, is crucial for a well-performing XGBoostPP. To this end, we propose a two-fold cross-validation method. We randomly split data points into two parts and, at each time, use one part to train the model and then compute the Poisson log-likelihood based on the estimated intensities on the other test part. We summarize the Poisson likelihoods on two test parts as the performance measure. To further reduce the randomness in data splitting, we repeat such a procedure three times and report the averaged test Poisson log-likelihoods as the final evaluation metric for choosing the optimal combination of hyperparameters.

\section{Simulation Study}
\label{sec:simulation}

To evaluate the performance of XGBoostPP, we conduct a simulation study on synthetic data generated from the point process models introduced in Sec.~\ref{subsec:PPmodels}. The intensity is governed by a set of covariates that are independently generated from an isotropic Gaussian process with a covariance function $\exp(-10r)$. For comparison, we adopt KIEs with the ratio (ra) and the reweight (re) methods \citep{Baddeley2012KIE} and select the bandwidths by Silverman's rule of thumb \citep{Silverman2018Book}. It is again worth noting that KIEs implemented in the R-package \textit{spatstat} only support data with at most two covariates. For GCPs, we implement APP with the naive kernel (naive) and the degenerate kernel of the random Fourier map (rfm) \citep{Kim2022AugmentedPP}. We find that the naive kernel also encounters computation problems when estimating the intensity function over a high-dimensional covariate space. 

We use two evaluation metrics to quantitatively measure the performance of XGBoostPP and other approaches: the Poisson log-likelihood on new randomly generated test data under the same point process model (reported in the main text) and the integrated absolute error from true intensities (reported in Appendix~\ref{sec:Appendix_C}). For each scenario, we report the averaged values and include the standard deviations of the two metrics across $500$ simulation runs. Note that, in all tables throughout this paper, the ‘XGBoostPP’ terms with subscripts ‘p’ and ‘wp’ represent the models fitted using the Poisson likelihood loss function and the weighted Poisson likelihood loss function, respectively.

In all simulation runs, the hyperparameters of the XGBoostPP model -- $K$, $\gamma$ and $\eta$ -- are chosen through the two-fold cross-validation method in Sec.~\ref{subsec:hyperparameterselection} from the candidate sets $\{1, \dots, 600\}$, $\{10, 30, 50\}$ and $\{0.1, 0.05, 0.01\}$. Moreover, we employ ten parallel trees for each training iteration of XGBoostPP and set the proportion of the subsampled covariates at a node split to $1/3$.

\begin{table}[ht]
\caption{Averaged Poisson log-likelihoods (standard deviations) of different approaches on Poisson test data.}
\vskip -0.1in
\label{tab:1}
\begin{center}
\begin{scriptsize}
\begin{sc}
\setlength{\tabcolsep}{4pt}
\begin{tabular}{p{0.8cm}p{2cm}ll}
\toprule
{\centering Covs} & Poisson process & $\beta=0.5$ & $\beta=1.0$\\
\midrule
& $\mathrm{True}$  & $2089.4$ & $2290.3$\\
\cmidrule(lr){3-4}
\multirow{6}{*}{\parbox{0.8cm} {\centering $2$\\ \centering (§4.1)}} & $\mathrm{KIE}_{\mathrm{ra}}$ & $2072.9(3.2)$ & $2226.6(5.9)$\\
& $\mathrm{KIE}_{\mathrm{re}}$ & $2049.6(3.9)$ & $2184.6(6.6)$\\
& $\mathrm{APP}_{\mathrm{naive}}$ & $2077.7(3.9)$ & $\pmb{2273.6}(4.3)$\\
& $\mathrm{APP}_{\mathrm{rfm}}$ & $\pmb{2077.8}(3.9)$ & $2273.5(4.2)$\\
& $\mathrm{XGBoostPP}_{\mathrm{p}}$ & $2073.2(3.0)$ & $2261.4(5.0)$\\
& $\mathrm{XGBoostPP}_{\mathrm{wp}}$& $2073.2(3.0)$ & $2261.6(5.3)$\\
\cmidrule(lr){3-4}
\multirow{3}{*}{\parbox{0.8cm} {\centering $10$\\ \centering (§4.3)}}& $\mathrm{APP}_{\mathrm{rfm}}$ & $1943.8(15.0)$ & $2127.3(14.4)$\\
& $\mathrm{XGBoostPP}_{\mathrm{p}}$ & $2062.3(4.5)$ & $\pmb{2245.4}(5.5)$\\
& $\mathrm{XGBoostPP}_{\mathrm{wp}}$& $\pmb{2062.4}(4.5)$ & $\pmb{2245.4}(5.5)$\\
\midrule
Covs & Poisson process & $\beta=0.2$ & $\beta=0.4$\\
\midrule
& $\mathrm{True}$ & $2048.1$ & $2160.9$\\
\cmidrule(lr){3-4}
\multirow{3}{*}{\parbox{0.8cm} {\centering $10$\\ \centering (§4.2)}} & $\mathrm{APP}_{\mathrm{rfm}}$ & $1857.9(41.2)$ & $1962.2(18.0)$\\
& $\mathrm{XGBoostPP}_{\mathrm{p}}$ & $\pmb{2025.9}(3.8)$ & $\pmb{2115.9}(6.3)$\\
& $\mathrm{XGBoostPP}_{\mathrm{wp}}$& $\pmb{2025.9}(3.8)$ & $\pmb{2115.9}(6.3)$\\
\bottomrule
\end{tabular}
\end{sc}
\end{scriptsize}
\end{center}
\vskip -0.2in
\end{table}

\begin{table*}[ht]
\caption{Averaged Poisson log-likelihoods (standard deviations) of different approaches on log-Gaussian Cox test data.}
\label{tab:2}
\vskip 0.05in
\begin{center}
\begin{scriptsize}
\begin{sc}
\setlength{\tabcolsep}{4pt}
\begin{tabular}{p{0.8cm}p{2cm}llll}
\toprule
\multirow{2}{*}{\parbox{0.8cm} {\centering Covs}} & \multirow{2}{*}{\parbox{1.8cm} {Log-Gaussian\\ Cox process}} & \multicolumn{2}{c}{$\tau^2=1$} &\multicolumn{2}{c}{$\tau^2=2$}\\
& & $\sigma=0.02$ & $\sigma=0.04$ & $\sigma=0.02$ & $\sigma=0.04$\\
\midrule
& & \multicolumn{4}{c}{$\beta=0.5$}\\ \cmidrule(lr){3-6}
& $\mathrm{True}$ & $2070.2$ & $2080.5$ & $2089.7$ & $2102.5$\\
\cmidrule(lr){3-6}
\multirow{6}{*}{\parbox{0.8cm} {\centering $2$\\ \centering (§4.1)}} & $\mathrm{KIE}_{\mathrm{ra}}$ & $2052.7(5.4)$ & $2060.5(7.8)$ & $2071.5(7.1)$ & $2077.4(12.8)$\\
& $\mathrm{KIE}_{\mathrm{re}}$    & $2029.4(6.5)$ & $2037.1(8.7)$ & $2049.0(8.2)$ & $2051.8(14.1)$\\
& $\mathrm{APP}_{\mathrm{naive}}$   & $2057.2(6.6)$ & $2064.3(9.3)$ & $2074.5(8.2)$ & $2075.7(17.1)$\\
& $\mathrm{APP}_{\mathrm{rfm}}$   & $\pmb{2057.3}(6.6)$ & $\pmb{2064.6}(9.2)$ & $\pmb{2074.8}(8.2)$ & $2076.2(16.7)$\\
& $\mathrm{XGBoostPP}_{\mathrm{p}}$ & $2051.8(5.0)$ & $2059.0(8.0)$ & $2069.0(6.9)$ & $2074.0(14.7)$\\
& $\mathrm{XGBoostPP}_{\mathrm{wp}}$ & $2052.6(5.7)$ & $2060.9(7.7)$ & $2072.1(6.8)$ & $\pmb{2078.9}(12.9)$\\
\cmidrule(lr){3-6}
\multirow{3}{*}{\parbox{0.8cm} {\centering $10$\\ \centering (§4.3)}} & $\mathrm{APP}_{\mathrm{rfm}}$   & $1909.3(21.7)$ & $1901.5(28.9)$ & $1907.1(28.3)$ & $1872.4(42.0)$\\
& $\mathrm{XGBoostPP}_{\mathrm{p}}$ & $2042.5(6.5)$ & $2049.4(8.8)$ & $2059.0(11.0)$ & $2054.5(25.0)$\\
& $\mathrm{XGBoostPP}_{\mathrm{wp}}$ & $\pmb{2044.2}(6.9)$ & $\pmb{2051.6}(9.0)$ & $\pmb{2062.7}(8.4)$ & $\pmb{2061.5}(16.3)$\\
\cmidrule(lr){3-6}
& & \multicolumn{4}{c}{$\beta=1.0$}\\ \cmidrule(lr){3-6}
& $\mathrm{True}$ &$2304.8$ & $2269.1$ & $2256.0$ & $2283.2$\\
\cmidrule(lr){3-6}
\multirow{6}{*}{\parbox{0.8cm} {\centering $2$\\ \centering (§4.1)}} & $\mathrm{KIE}_{\mathrm{ra}}$ & $2238.8(11.4)$ & $2202.3(15.7)$ & $2188.2(15,1)$ & $2211.1(22.8)$\\
& $\mathrm{KIE}_{\mathrm{re}}$    & $2196.2(12.5)$ & $2159.4(16.2)$ & $2144.2(16.5)$ & $2166.7(23.5)$\\
& $\mathrm{APP}_{\mathrm{naive}}$   & $2284.7(8.2)$ & $2244.0(12.6)$ & $2229.6(13.4)$ & $2243.4(27.7)$\\
& $\mathrm{APP}_{\mathrm{rfm}}$   & $\pmb{2284.7}(8.1)$ & $\pmb{2244.0}(12.5)$ & $\pmb{2229.9}(12.8)$ & $2244.4(25.9)$\\
& $\mathrm{XGBoostPP}_{\mathrm{p}}$ & $2272.2(8.0)$ & $2231.8(11.8)$ & $2219.4(11.6)$ & $2234.5(21.6)$\\
& $\mathrm{XGBoostPP}_{\mathrm{wp}}$ & $2275.8(8.7)$ & $2235.5(12.9)$ & $2228.8(9.6)$ & $\pmb{2244.9}(18.6)$\\
\cmidrule(lr){3-6}
\multirow{3}{*}{\parbox{0.8cm} {\centering $10$\\ \centering (§4.3)}} & $\mathrm{APP}_{\mathrm{rfm}}$   & $2116.5(22.6)$ & $2062.9(30.7)$ & $2038.5(31.7)$ & $2021.0(51.3)$\\
& $\mathrm{XGBoostPP}_{\mathrm{p}}$ & $2256.4(8.3)$ & $2214.6(12.6)$ & $2199.3(14.7)$ & $2207.3(32.8)$\\
& $\mathrm{XGBoostPP}_{\mathrm{wp}}$ & $\pmb{2258.8}(9.1)$ & $\pmb{2215.7}(13.0)$ & $\pmb{2208.5}(12.2)$ & $\pmb{2215.1}(23.5)$\\
\midrule
& & \multicolumn{4}{c}{$\beta=0.2$}\\ \cmidrule(lr){3-6} 
& $\mathrm{True}$ & $2032.9$ & $2066.7$ & $2080.8$ & $2080.7$\\
\cmidrule(lr){3-6}
\multirow{3}{*}{\parbox{0.8cm} {\centering $10$\\ \centering (§4.2)}} & $\mathrm{APP}_{\mathrm{rfm}}$   & $1848.3(31.6)$ & $1871.8(30.8)$ & $1879.9(34.4)$ & $1838.5(42.0)$\\
& $\mathrm{XGBoostPP}_{\mathrm{p}}$   & $2011.5(4.3)$ & $2041.7(7.0)$ & $2054.1(11.1)$ & $2040.5(19.0)$\\
& $\mathrm{XGBoostPP}_{\mathrm{wp}}$ & $\pmb{2012.8}(4.2)$ & $\pmb{2042.4}(7.0)$ & $\pmb{2056.0}(7.8)$ & $\pmb{2042.3}(15.0)$\\ \cmidrule(lr){3-6}
& & \multicolumn{4}{c}{$\beta=0.4$}\\ \cmidrule(lr){3-6}
& $\mathrm{True}$ & $2134.6$ & $2181.9$ & $2153.3$ & $2171.0$\\
\cmidrule(lr){3-6}
\multirow{3}{*}{\parbox{0.8cm} {\centering $10$\\ \centering (§4.2)}} & $\mathrm{APP}_{\mathrm{rfm}}$   & $1923.8(24.5)$ & $1944.8(30.6)$ & $1916.3(30.3)$ & $1887.6(45.7)$\\
& $\mathrm{XGBoostPP}_{\mathrm{p}}$   & $2092.0(7.8)$ & $2130.7(12.4)$ & $2106.2(12.4)$ & $2105.4(24.6)$\\
& $\mathrm{XGBoostPP}_{\mathrm{wp}}$ & $\pmb{2096.0}(8.7)$ & $\pmb{2133.5}(11.8)$ & $\pmb{2112.6}(10.8)$ & $\pmb{2110.6}(20.3)$\\
\bottomrule
\end{tabular}
\end{sc}
\end{scriptsize}
\end{center}
\vskip -0.2in
\end{table*}

\begin{table}[ht]
\caption{Averaged Poisson log-likelihoods (standard deviations) of different approaches on Neyman-Scott test data.}
\label{tab:3}
\vskip -0.5in
\begin{center}
\begin{scriptsize}
\begin{sc}
\setlength{\tabcolsep}{4pt}
\begin{tabular}{p{0.8cm}p{2cm}llll}
\toprule
\multirow{2}{*}{\parbox{0.8cm} {Covs}} & \multirow{2}{*}{\parbox{2cm} {Neyman--Scott\\ process}} & \multicolumn{2}{c}{$\kappa=100$} & \multicolumn{2}{c}{$\kappa=200$}\\
& & $\sigma=0.02$ & $\sigma=0.04$ & $\sigma=0.02$ & $\sigma=0.04$\\ 
\midrule
& & \multicolumn{4}{c}{$\beta=0.5$}\\ \cmidrule(lr){3-6}
& $\mathrm{True}$ & $2090.9$ & $2074.8$ & $2091.2$ & $2071.7$\\
\cmidrule(lr){3-6}
\multirow{6}{*}{\parbox{0.8cm} {\centering $2$\\ \centering (§4.1)}} & $\mathrm{KIE}_{\mathrm{ra}}$ & $2071.1(8.0)$ & $2055.2(6.9)$ & $2072.2(6.2)$ & $2053.4(5.1)$\\
& $\mathrm{KIE}_{\mathrm{re}}$    & $2047.6(10.0)$ & $2031.0(8.0)$ & $2049.7(7.4)$ & $2030.4(6.0)$\\
& $\mathrm{APP}_{\mathrm{naive}}$   & $2073.3(10.4)$ & $2060.2(8.0)$ & $2077.5(7.2)$ & $2059.0(5.9)$\\
& $\mathrm{APP}_{\mathrm{rfm}}$   & $\pmb{2073.6}(10.2)$ & $\pmb{2060.4}(8.0)$ & $\pmb{2077.6}(7.2)$ & $\pmb{2059.1}(5.8)$\\
& $\mathrm{XGBoostPP}_{\mathrm{p}}$ & $2068.0(8.0)$ & $2054.6(6.9)$ & $2070.7(6.1)$ & $2052.8(4.8)$\\
& $\mathrm{XGBoostPP}_{\mathrm{wp}}$ & $2071.2(8.1)$ & $2056.2(7.7)$ & $2072.6(6.7)$ & $2053.1(5.0)$\\
\cmidrule(lr){3-6}
\multirow{3}{*}{\parbox{0.8cm} {\centering $10$\\ \centering (§4.3)}} & $\mathrm{APP}_{\mathrm{rfm}}$   & $1890.8(34.2)$ & $1905.0(26.5)$ & $1919.6(24.6)$ & $1912.1(21.0)$\\
& $\mathrm{XGBoostPP}_{\mathrm{p}}$ & $2057.6(11.5)$ & $2044.4(8.4)$ & $2061.4(7.3)$ & $2043.0(6.2)$\\
& $\mathrm{XGBoostPP}_{\mathrm{wp}}$ & $\pmb{2060.9}(9.8)$ & $\pmb{2046.4}(8.3)$ & $\pmb{2064.0}(7.7)$ & $\pmb{2043.2}(6.3)$\\ \cmidrule(lr){3-6}
& & \multicolumn{4}{c}{$\beta=1.0$} \\ \cmidrule(lr){3-6}
& $\mathrm{True}$ & $2263.1$ & $2255.6$ & $2272.5$ & $2306.9$\\
\cmidrule(lr){3-6}
\multirow{6}{*}{\parbox{0.8cm} {\centering $2$\\ \centering (§4.1)}} & $\mathrm{KIE}_{\mathrm{ra}}$ & $2195.3(17.5)$ & $2190.0(14.2)$ & $2206.2(13.5)$ & $2238.9(10.8)$\\
& $\mathrm{KIE}_{\mathrm{re}}$    & $2151.6(19.5)$ & $2148.3(14.8)$ & $2162.8(14.8)$ & $2194.9(10.9)$\\
& $\mathrm{APP}_{\mathrm{naive}}$   & $2233.6(16.0)$ & $2233.5(11.8)$ & $\pmb{2248.9}(11.7)$ & $2285.6(8.1)$\\
& $\mathrm{APP}_{\mathrm{rfm}}$   & $\pmb{2234.4}(15.1)$ & $\pmb{2233.6}(11.6)$ & $2248.9(11.9)$ & $\pmb{2285.8}(8.1)$\\
& $\mathrm{XGBoostPP}_{\mathrm{p}}$ & $2222.1(13.9)$ & $2219.9(10.8)$ & $2236.4(10.6)$ & $2273.2(7.7)$\\
& $\mathrm{XGBoostPP}_{\mathrm{wp}}$ & $2234.1(12.6)$ & $2223.7(11.2)$ & $2246.2(9.5)$ & $2274.2(8.4)$\\
\cmidrule(lr){3-6}
\multirow{3}{*}{\parbox{0.8cm} {\centering $10$\\ \centering (§4.3)}} & $\mathrm{APP}_{\mathrm{rfm}}$   & $2022.2(39.8)$ & $2054.3(31.8)$ & $2063.8(31.7)$ & $2120.1(24.9)$\\
& $\mathrm{XGBoostPP}_{\mathrm{p}}$ & $2202.1(17.2)$ & $2203.3(11.8)$ & $2220.9(11.4)$ & $2257.2(8.5)$\\
& $\mathrm{XGBoostPP}_{\mathrm{wp}}$ & $\pmb{2211.7}(19.9)$ & $\pmb{2204.6}(11.9)$ & $\pmb{2228.7}(11.9)$ & $\pmb{2257.2}(8.5)$\\
\midrule
& & \multicolumn{4}{c}{$\beta=0.2$}\\ \cmidrule(lr){3-6}
& $\mathrm{True}$ & $1977.8$ & $1963.7$ & $2070.6$ & $2047.9$\\
\cmidrule(lr){3-6}
\multirow{3}{*}{\parbox{0.8cm} {\centering $10$\\ \centering (§4.2)}} & $\mathrm{APP}_{\mathrm{rfm}}$   & $1777.1(33.2)$ & $1786.8(27.5)$ & $1884.5(27.5)$ & $1865.6(29.5)$\\
& $\mathrm{XGBoostPP}_{\mathrm{p}}$ & $1950.2(10.3)$ & $1938.4(6.3)$ & $2048.1(5.1)$ & $2025.8(4.8)$\\
& $\mathrm{XGBoostPP}_{\mathrm{wp}}$ & $\pmb{1952.1}(8.1)$ & $\pmb{1939.8}(6.1)$ & $\pmb{2049.2}(5.4)$ & $\pmb{2026.1}(4.7)$\\ \cmidrule(lr){3-6}
& & \multicolumn{4}{c}{$\beta=0.4$}\\ \cmidrule(lr){3-6}
& $\mathrm{True}$ & $2227.0$ & $2214.4$ & $2128.7$ & $2175.2$\\
\cmidrule(lr){3-6}
\multirow{3}{*}{\parbox{0.8cm} {\centering $10$\\ \centering (§4.2)}} & $\mathrm{APP}_{\mathrm{rfm}}$  & $1955.6(39.4)$ & $1980.7(32.3)$ & $1916.2(30.2)$ & $1961.8(24.3)$\\
& $\mathrm{XGBoostPP}_{\mathrm{p}}$ & $2173.2(13.0)$ & $2163.9(12.0)$ & $2087.4(8.1)$ & $2128.3(8.4)$\\
& $\mathrm{XGBoostPP}_{\mathrm{wp}}$ & $\pmb{2176.6}(16.0)$ & $\pmb{2166.0}(11.8)$ & $\pmb{2089.0}(8.4)$ & $\pmb{2128.4}(8.4)$\\
\bottomrule
\end{tabular}
\end{sc}
\end{scriptsize}
\end{center}
\vskip -0.2in
\end{table}

\subsection{Low-dimensional Covariate Space}
\label{subsec:lowdimensionaldata}
% Note use of \abovespace and \belowspace to get reasonable spacing
% above and below tabular lines.

In this scenario, the intensity function of the point process is determined by two covariates, denoted by $z_1(\bm{s})$ and $z_2(\bm{s})$. For the Poisson process, we set $\lambda(\bm s)=\alpha\exp[\beta z_1(\bm{s})+\beta z_2(\bm{s})]$, where $\beta=0.5$ or $1.0$ and $\alpha$ is chosen such that the expected number of events in $\mathcal{S}=[0,1]^2$ is $400$. For the log-Gaussian Cox process, we simulate point pattern data from a Poisson process with the random intensity $\Lambda(\bm s)=\alpha\exp[\beta z_1(\bm{s})+\beta z_2(\bm{s})+Y(\bm{s})]$, where $Y(\bm{s})$ is a zero-mean isotropic Gaussian random field with a covariance function $\tau^2\exp(-r/\sigma)$ and $\tau^2=1.0$ or $2.0$ and $\sigma=0.02$ or $0.04$. For the Neyman-Scott process, we generate the parents from a homogeneous Poisson process with intensity $\kappa=100$ or $200$ and, for each parent, simulate the offspring points using a Gaussian kernel with a standard deviation $\sigma=0.02$ or $0.04$. Following \citet{Guan2010WeightedPoissonLoss}, the distance $m$ used to approximate the weights \eqref{e:6} is set to $0.06$ for the Poisson process and to $3\sigma$ for the other two processes. 

%simulate a Poisson number of offspring points, $\max_{\bm{s}\in \mathcal{S}}\{\alpha\exp[\beta z_1(\bm{s})+\beta z_2(\bm{s})]\}$, where the location of each relative to its parent follows a Gaussian distribution with standard deviation $\sigma=0.02$ or $0.04$. We then keep every offspring point with a probability $\exp[\beta z_1(\bm{s})+\beta z_2(\bm{s})]/\max_{\bm{s}\in \mathcal{S}}\{\exp[\beta z_1(\bm{s})+\beta z_2(\bm{s})]\}$. Following \citet{Guan2010WeightedPoissonLoss}, the distance $m$ used for the spatial weights (\ref{e:6}) is set to $0.06$ for the Poisson process and to $3\sigma$ for the other two processes. 

We report the averaged Poisson log-likelihood results on Poisson test data in Tab.~\ref{tab:1} and those on log-Gaussian Cox and Neyman-Scott test data in Tab.~\ref{tab:2} and~\ref{tab:3}. In general, when there are only two covariates, XGBoostPP achieves competitive performance on all types of data under various parameter settings, although APP mostly exhibits the best behavior. For all approaches, intensity estimation is challenging when observed point patterns are more spatially varying (i.e.,\ large $\beta$) and clustered (i.e.,\ large $\tau^2$ and $\sigma$ for the log-Gaussian Cox process and small $\kappa$ and $\sigma$ for the Neyman-Scott process). Moreover, comparing XGBoostPP$_{\rm p}$ with XGBoostPP$_{\rm wp}$, it is evident that the performance improves on clustered processes while it remains almost unchanged on Poisson processes. This indicates that the practical training algorithm designed in Sec.~\ref{subsec:TrainingManner} works well when the clustering feature of a point pattern is unknown. More specifically, these improvements are greater when point patterns are more spatially heterogeneous and clustered, aligning with the findings in \citet{Guan2010WeightedPoissonLoss}.

\subsection{Higher-dimensional Covariate Space}
\label{subsec:highdimensionaldata}

In this scenario, we conduct similar experiments but increase the number of accessible covariates to ten. Denote the covariates by $z_1(\bm{s}), \dots, z_{10}(\bm{s})$. The intensity function of the point process is now defined as $\lambda(\bm s)=\alpha\exp(\beta \{z_1(\bm{s})+\frac{1}{2}z_2(\bm{s})z_3(\bm{s})+ \frac{1}{6}\exp[z_4(\bm{s})]+ \frac{1}{2}z_5(\bm{s})^2 + 3\sin[z_6(\bm{s})]\})$. Note that $z_7(\bm{s}),\dots, z_{10}(\bm{s})$ are employed as nuisance variables which indeed exist often in real-world applications. We set $\sigma,\tau^2,\kappa$ to the same values as in Sec.~\ref{subsec:lowdimensionaldata} while $\beta=0.2$ or $0.4$ to balance the spatial heterogeneity of point patterns to a reasonable scale. Considering that the detection of clustering patterns over complex covariate relationships is difficult, we change $m$ to $0.04$ for the Poisson process and to $2\sigma$ for the log-Gaussian Cox and the Neyman-Scott processes.

The relevant evaluation results are also reported in Tab.~\ref{tab:1}, ~\ref{tab:2} and ~\ref{tab:3}. When there are ten covariates, the performance of APP, with the `rfm' kernel, deteriorates considerably, while XGBoostPP maintains good performance. Moreover, the standard deviations given by APP tend to be larger than those by XGBoostPP, indicating a more stable estimation accuracy for XGBoostPP. Comparing XGBoostPP$_{\rm p}$ with XGBoostPP$_{\rm wp}$, the latter again shows better performance and the improvements are, overall, slightly higher for more spatially heterogeneous and clustered point patterns. However, compared to those in the lower-dimensional case, such improvements become less distinguished.

\subsection{Simple Intensity with Many Nuisance Covariates}
\label{subsec:additionalsimulationstudy}

In this section, we simulate data of the same models as those in Sec.~\ref{subsec:lowdimensionaldata}, using the simple function $\lambda(\bm s)=\alpha\exp[\beta z_1(\bm{s})+\beta z_2(\bm{s})]$. However, we input the ten covariates from Sec.~\ref{subsec:highdimensionaldata}, $z_1(\bm{s}),\ldots,z_{10}(\bm{s})$, to estimate the intensity, thus with eight nuisance covariates. Such a scenario is common in practice when numerous covariates are available for point process intensity estimation while researchers are uncertain about which ones are most relevant. We again evaluate XGBoostPP and APP with the two performance metrics, the Poisson log-likelihood on new generated data and the integrated absolute error, and report the relevant results. From Tab.~\ref{tab:1}, ~\ref{tab:2} and ~\ref{tab:3}, it is obvious that XGBoostPP exhibits much more robust behavior in the presence of many nuisance covariates compared to APP, which can be evidenced by the significantly smaller reductions in the test Poisson log-likelihoods. Such robustness suggests great non-parametric potential of XGBoostPP in practical applications.

\subsection{Poisson Toy Examples}

In addition to the main simulation study, we test XGBoostPP on three Poisson toy examples in Appendix~\ref{sec:Appendix_D} to intuitively visualize its flexibility in modeling various nonlinear relationships between covariates and the intensity function.

\section{Real Data Analyses}
\label{sec:realworldStudy}

To demonstrate the utility of XGBoostPP in practice, we apply it to two real data sets: tropical forest data on Barro Colorado Island (BCI) in Panama and kitchen fire data in the Twente region in The Netherlands. For every data set, we conduct a four-fold cross-validation to evaluate different intensity estimation approaches based on the test Poisson log-likelihood (i.e.,\ randomly splitting the data points into four subsets, assigning one for test and the others for training). Details on the covariate information in the two data sets and the calculation of the test Poisson log-likelihood metric used here are given in Appendix~\ref{sec:Appendix_E1} and \ref{sec:Appendix_E2}.

\begin{table}[ht]
\caption{Cross-validated (four folds) test Poisson log-likelihoods of different approaches on ‘Bei’, ‘Capp’ and ‘Fire’ data.}
\label{tab:4}
\vskip -0.4in
\begin{center}
\begin{scriptsize}
\begin{sc}
\setlength{\tabcolsep}{6pt}
\begin{tabular}{p{1.9cm}ccc}
\toprule
Data set & Bei & Capp & Fire\\ \cmidrule(lr){2-4}
Covs & $8$ & $8$ & $29$\\
\midrule
$\mathrm{APP}_{\mathrm{rfm}}$ & $-24955.4$ & $-27900.1$ & -\\
$\mathrm{XGBoostPP}_{\mathrm{p}}$ & $-24557.6$ & $-27701.7$ & $-10773.2$\\
$\mathrm{XGBoostPP}_{\mathrm{wp}}$ & $\pmb{-24376.1}$ & $\pmb{-27684.9}$ & $\pmb{-10519.7}$\\
\bottomrule
\end{tabular}
\end{sc}
\end{scriptsize}
\end{center}
\vskip -0.1in
\end{table}

\begin{figure*}[ht]
\begin{center}
\centerline{\includegraphics[scale=0.37]{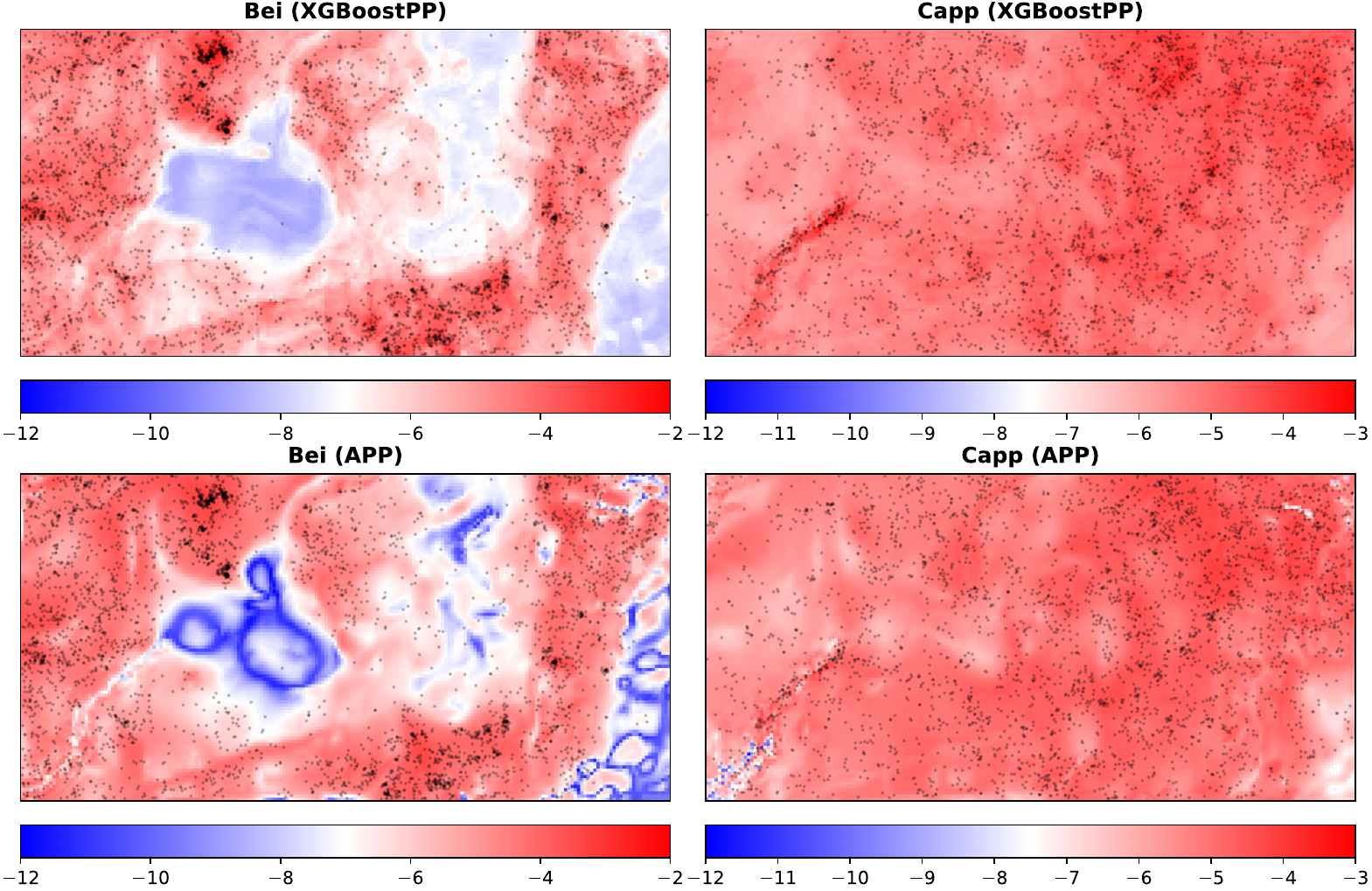}}
\caption{Observed tree locations and estimated log-intensities by XGBoostPP$_{\rm wp}$ and APP over 8 covariates.}
\label{fig:1}
\end{center}
\vskip -0.1in
\end{figure*}

\textbf{BCI Forestry Data.} From the BCI forestry data, we select two tree species to study: Beilschmiedia pendula (Bei) and Capparis frondosa (Capp), containing $3604$ and $3299$ tree locations, respectively. We investigate eight covariates that may contribute to explaining their spatial distributions, including terrain elevation and slope, four soil nutrients and solar and wetness indices. Proper intensity estimation for tree species helps forestry scientists research their suitable living environments. Tab.\ref{tab:4} indicates that, on both `Bei' and `Capp' data, XGBoostPP outperforms APP. Since `Bei' appears more clustered compared to `Capp' (see, \citealp{Guan2010WeightedPoissonLoss,Yue2011GCPMcmc}), XGBoostPP$_{\rm wp}$ improves the performance on the former relatively more. Fig.\ref{fig:1} displays the estimated log-intensities, showing that, in comparison to APP, XGBoostPP produces significantly different estimates for the areas with a small number of event observations.

\begin{figure}[ht]
\begin{center}
\centerline{\includegraphics[scale=0.47]{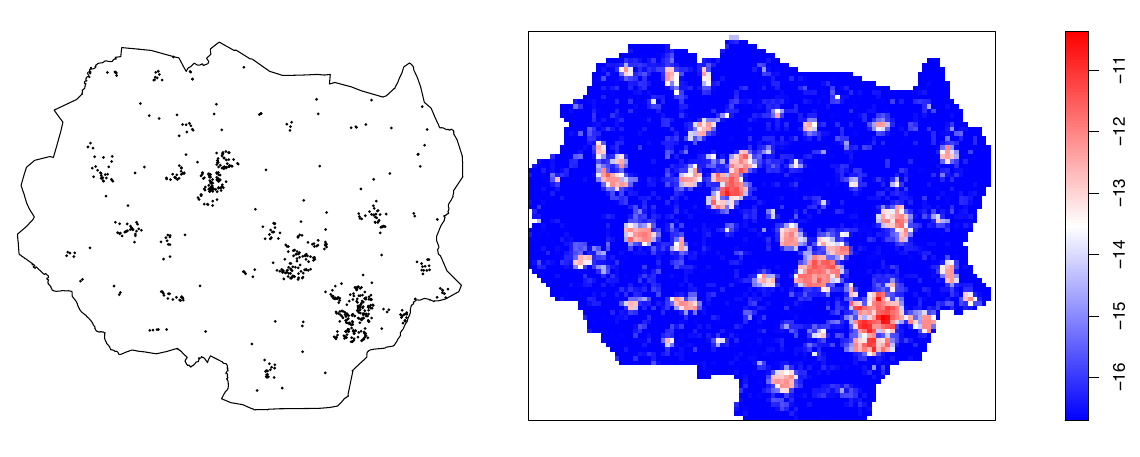}}
\caption{Kitchen fire locations and estimated log-intensity by XGBoostPP$_{\rm wp}$ over 29 covariates.}
\label{fig:2}
\end{center}
\vskip -0.1in
\end{figure}

\textbf{Kitchen Fire Data.} The fire data (Fire) comprises $699$ kitchen fire locations in Twente from $2004$ to $2020$. The covariates of interest include building information, urbanity degrees, population components, and energy consumption ($29$ covariates in total). Identifying the most relevant covariates for fire occurrences enables firefighters to organize public campaigns for fire prevention more effectively. Tab.\ref{tab:4} provides estimation results for XGBoostPP only, as APP reports computational errors when fitted with 29 covariates. On `Fire' data, XGBoostPP demonstrates its strong capacity to handle covariate spaces of high dimensions. As expected, XGBoostPP$_{\rm wp}$ again yields a higher test Poisson log-likelihood due to the apparent spatial dependence observed in the raw data. Finally, the estimated log-intensity for `Fire' by XGBoostPP is presented in Fig.\ref{fig:2}.

%\subsection{Practical Utility}

%\vspace{0.5mm}
%The estimated intensities by XGBoostPP better reflects the distribution of point occurrences, which can help forestry scientists learn the suitable living environment for tree species and can provide firefighters with the risk information for proper personnel and equipment arrangements. Moreover, XGBoostPP can accommodate more available covariates, providing prior knowledge for further analyzing the relationship between a putative variable and the intensity function. For instance, forestry researcher may thus be motivated to investigate the specific influence of a nutrient on the growth of a tree specie and firefighters may use the information to organize public campaigns for fire prevention.

\section{Conclusions}

In this paper, we proposed a novel tree-based ensemble method, named XGBoostPP, to nonparametrically estimate the intensity of a point process as a function of spatial covariates. Two loss functions were carefully designed for model estimation. The first one is the Poisson likelihood loss, working for general point processes. However, when the underlying process is not Poisson, this loss can be inefficient. The second one is the weighted Poisson likelihood loss, where a weight function is introduced to account for potential spatial dependence in order to further improve the estimation efficiency for clustered processes. We developed an efficient greedy search algorithm for model fitting and proposed a two-fold cross-validation procedure to select hyperparameters.

To show the effectiveness of XGBoostPP, we performed extensive simulation studies on data over low- and higher-dimensional covariate spaces and demonstrated its advantages against existing approaches. In particular, XGBoostPP achieves superior performance when the dimension of the covariate space is large. The applications on the two real-world data sets reveal that a well-constructed and -tuned XGBoostPP can flexibly analyze unknown, complex point patterns in practice.

For future work, the first direction would be to improve the estimation of Ripley's K-function for point patterns with high-dimensional covariate spaces, so that the approximated weighted Poisson likelihood loss could be improved. Moreover, it would also be interesting to extend the proposed method to estimating second-order intensities for non-stationary point pattern data.

\bibliography{paper}
\bibliographystyle{neurips_2023}

\newpage
\appendix
\onecolumn

\counterwithin*{equation}{section}
\renewcommand\theequation{\thesection\arabic{equation}}

\section{Motivation for the Weighted Poisson Likelihood Loss in Sec.~\ref{subsec:LossFunctions}}
\label{sec:Appendix_A}

We motivate the weighted Poisson likelihood loss by carefully analyzing the estimation efficiency of the scores on the non-`terminal' node (true supporting leaves) under the `oracle' tree structures of XGBoostPP.

Note that, with the knowledge of the optimal tree structures $\bm{q}^{\mathcal{O}}$, the XGBoostPP model can be interpreted as a generalized linear model over a vector of transformed covariates $\bm{Z}_{\mathcal{A}}(\bm{s})$ that denote the membership identities of a location $\bm{s}$ with respect to the supporting tree leaves. Specfically, 
\begin{equation*}
    \lambda(\bm{s};\boldsymbol{\theta}_{\mathcal{A}})=\exp[\boldsymbol{\theta}_{\mathcal{A}}^{\top}\bm{Z}_{\mathcal{A}}(\bm{s})].
\end{equation*}
In this setting, the estimated intensity function is positive and differentiable with respect to $\boldsymbol{\theta}_{\mathcal{A}}$.

By the Karush–Kuhn–Tucker optimality condition, the estimate of $\bm\theta_{\mathcal{A}}$ obtained by minimizing
\begin{equation*}
    L(\bm f)=\sum_{k=1}^K\Omega(f_k)-\sum_{\bm{x}\in \bm{\mathrm{x}}}\log\left[\lambda(\bm{x};\bm f)\right]
    +\int_{\mathcal{S}}\lambda(\bm{s};\bm f)\mathrm{d}\bm{s},
\end{equation*}
must satisfy the estimating equation
\begin{equation*}
    \bm{e}_{\lambda}(\boldsymbol{\theta}_{\mathcal{A}})=\gamma \mathrm{sgn}(\boldsymbol{\theta}_{\mathcal{A}})+\sum_{\bm{x}\in \bm{\mathrm{x}}}\bm{Z}_{\mathcal{A}}(\bm{s})-\int_{\mathcal{S}} \lambda(\bm{s};\boldsymbol{\theta}_{\mathcal{A}})\bm{Z}_{\mathcal{A}}(\bm{s})\mathrm{d}\bm{s} = \bm{0},
\end{equation*}
which, according to \citet{Guan2010WeightedPoissonLoss}, in a more general form is 
\begin{equation}
    \bm{e}_{\bm{h}}(\boldsymbol{\theta}_{\mathcal{A}})=\gamma \mathrm{sgn}(\boldsymbol{\theta}_{\mathcal{A}})+\sum_{\bm{x}\in \bm{\mathrm{x}}}\bm{h}(\bm{s};\boldsymbol{\theta}_\mathcal{A})-\int_{\mathcal{S}} \lambda(\bm{s};\boldsymbol{\theta}_{\mathcal{A}})\bm{h}(\bm{s};\boldsymbol{\theta}_\mathcal{A})\mathrm{d}\bm{s} = \bm{0}.
    \label{e:A1}
\end{equation}
Here, $\bm{h}(\bm{s};\boldsymbol{\theta}_\mathcal{A})$ can be any measurable vector function of the same dimension as $\boldsymbol{\theta}_\mathcal{A}$. Following \citet{Guan2015Quasilikelihood}, we call \eqref{e:A1} a first-order penalized estimating function.

To improve the estimation efficiency of $\boldsymbol{\theta}_\mathcal{A}$ for general point processes, we analyze the estimation variance of \eqref{e:A1}. Recall that $\boldsymbol{\theta}_{\mathcal{A}}\neq \bm{0}$, and denote the sensitivity matrix
\begin{equation*}
    -\mathbb{E}\left\{\frac{\partial \bm{e}_{\bm{h}}(\boldsymbol{\theta}_{\mathcal{A}})}{\partial \boldsymbol{\theta}^{\top}_{\mathcal{A}}} \right\}
    =\int_{\mathcal{S}}\nabla\lambda(\bm{s};\boldsymbol{\theta}_{\mathcal{A}})\bm{h}^{\top}(\bm{s};\boldsymbol{\theta}_{\mathcal{A}})\mathrm{d}\bm{s}
\end{equation*}
by $\bm{S}_{\bm{h}}$ and the variance matrix 
\begin{equation*}
\begin{split}
    \mathrm{var}\left\{\bm{e}_{\bm{h}}(\boldsymbol{\theta}_{\mathcal{A}})\right\}
    &=\int_{\mathcal{S}}\bm{h}(\bm{s};\boldsymbol{\theta}_{\mathcal{A}})\bm{h}^{\top}(\bm{s};\boldsymbol{\theta}_{\mathcal{A}})\lambda(\bm{s};\boldsymbol{\theta}_{\mathcal{A}})\mathrm{d}\bm{s}\\
    &+\int_{\mathcal{S}}\int_{\mathcal{S}}\bm{h}(\bm{u};\boldsymbol{\theta}_{\mathcal{A}})\bm{h}^{\top}(\bm{s};\boldsymbol{\theta}_{\mathcal{A}})\lambda(\bm{u};\boldsymbol{\theta}_{\mathcal{A}})\lambda(\bm{s};\boldsymbol{\theta}_{\mathcal{A}})\left[g(\bm{s},\bm{u})-1\right]\mathrm{d}\bm{u}\mathrm{d}\bm{s}
\end{split}
\end{equation*}
by $\boldsymbol{\Sigma}_{\bm{h}}$. Under a proper scheme, the estimation covariance matrix reads $\bm{S}_{\bm{h}}^{-1}\boldsymbol{\Sigma}_{\bm{h}}(\bm{S}_{\bm{h}}^{-1})^{\top}$ and its inverse, $\bm{G}_{\bm{h}}=\bm{S}_{\bm{h}}^{\top}\boldsymbol{\Sigma}_{\bm{h}}^{-1}\bm{S}_{\bm{h}}$, is known as the Godambe information.

The optimal $\bm{h}^{*}(\bm{s};\boldsymbol{\theta}_{\mathcal{A}})$ that minimizes the inverse Godambe information reduces the estimation variance for $\boldsymbol{\theta}_{\mathcal{A}}$ the most. By \citet{Guan2015Quasilikelihood}, it must satisfy the following Fredholm integral equation of the second kind 
\begin{equation*}
    \int_{\mathcal{S}}\bm{h}^*(\bm{u};\boldsymbol{\theta}_{\mathcal{A}})\lambda(\bm{u};\boldsymbol{\theta}_{\mathcal{A}})\left[g(\bm{s},\bm{u})-1\right]\mathrm{d}\bm{u}+\bm{h}^*(\bm{s};\boldsymbol{\theta}_{\mathcal{A}})=\nabla\lambda(\bm{s};\boldsymbol{\theta}_{\mathcal{A}})/\lambda(\bm{s};\boldsymbol{\theta}_{\mathcal{A}}).
\end{equation*}
In the spirit of \citet{Guan2010WeightedPoissonLoss}, we assume that $\bm{h}^*(\bm{u};\boldsymbol{\theta}_{\mathcal{A}})\lambda(\bm{u};\boldsymbol{\theta}_{\mathcal{A}})\simeq \bm{h}^*(\bm{s};\boldsymbol{\theta}_{\mathcal{A}})\lambda(\bm{s};\boldsymbol{\theta}_{\mathcal{A}})$ for proximate pairs of $\bm{s}, \bm{u}$ and $g(\bm{s},\bm{u})\simeq 1$ for distant pairs, which yields a closed-form solution $\bm{h}^*(\bm{s};\boldsymbol{\theta}_{\mathcal{A}})=w^*(\bm{s};\bm\theta_{\mathcal{A}})\nabla\lambda(\bm{s};\boldsymbol{\theta}_{\mathcal{A}})/\lambda(\bm{s};\boldsymbol{\theta}_{\mathcal{A}})$  with a spatially dependent weight function
\begin{equation*}
    w^*(\bm{s};\boldsymbol{\theta}_{\mathcal{A}})=\frac{1}{1+\lambda(\bm{s};\boldsymbol{\theta}_{\mathcal{A}})\int_{\mathcal{S}}\left[g(\bm{s},\bm{u})-1\right]\mathrm{d}\bm{u}}.
\end{equation*}
It has the advantage that when the point process is second-order intensity-reweighted stationary and the spatial correlations in it vanish beyond a distance of $m$, $\int_{\mathcal{S}}\left[g(\bm{s},\bm{u})-1\right]\mathrm{d}\bm{u}=\int_{\mathcal{S}}\left[g(\bm{s}-\bm{u})-1\right]\mathrm{d}\bm{u}=\int_{0}^m2\pi r[g(r)-1]\mathrm{d}r$ can be approximated using Ripley’s K-function (see, e.g., \citealp{Lieshout2019SpatialStatistics}) by $K(m)-\pi m^2$.

\section{Derivation of the Additive Training Algorithm in Sec.~\ref{subsec:TrainingManner}}
\label{sec:Appendix_B}

Recalling \eqref{e:4}, we minimize the following penalized weighted Poisson likelihood loss function
\begin{equation*}
    L_w(\bm{f})=\sum_{k=1}^K\Omega(f_k)-\sum_{\bm{x}\in \bm{\mathrm{x}}}w(\bm{x};\bm{f})\phi(\bm{x};\bm{f})+\int_{\mathcal{S}}w(\bm{s};\bm{f})\exp\left[\phi(\bm{s};\bm{f})\right]\mathrm{d}\bm{s}
\end{equation*}
with $\Omega(f_k)=\gamma\sum_v |\theta_{kv}|$. Denote the estimated log-intensity over $k-1$ trees $\hat{\bm f}_{k-1}=\{\hat f_1,\cdots,\hat f_{k-1}\}$ by $\hat{\phi}(\bm{s};\hat{\bm f}_{k-1})$ with $\bm{s}\in \mathcal{S}$. At the $k$-th iteration, we add a tree predictor $\hat f_k$ to minimize
\begin{equation*}
    L_w^{(k)}(f_k)
    =\Omega(f_k)-\sum_{\bm{x}\in \bm{\mathrm{x}}}w(\bm{x};\bm{f})\left[\hat{\phi}(\bm{x};\hat{\bm f}_{k-1})+f_k(\bm{x})\right]+\int_{\mathcal{S}}w(\bm{s};\bm{f})\exp\left[\hat{\phi}(\bm{s};\hat{\bm f}_{k-1})+f_k(\bm{s})\right]\mathrm{d}\bm{s}.
\end{equation*}
Without the knowledge of the tree structure $q_k$ of $f_k$, we use the following iteratively updated spatial weight to better approximate $w(\bm{x};\bm{f})$,
\begin{equation*}
    \hat{w}_k(\bm{s})=\frac{\omega_k}{1+\exp\left[\hat{\phi}(\bm{s};\hat{\bm f}_{k-1})\right]\left[\hat{K}(m)-\pi m^2\right]}.
\end{equation*}

%Note the following technicalities
%\begin{equation*}
%    \frac{\partial L_w}{\partial \phi(\bm{x};\hat{\bm{f}}_{k-1}))}=-\hat{w}_{k}(\bm{x}),\quad
%   \frac{\partial^2 L_w}{[\partial \phi(\bm{x};\hat{\bm{f}}_{k-1})]^2}=0,\quad
%    \frac{\partial L_w}{\partial \phi(\bm{s})}=\hat{w}_{k}(\bm{s})\exp\left[\phi(\bm{s})\right],\quad
%   \frac{\partial^2 L_w}{[\partial \phi(\bm{s})]^2}=
%   \hat{w}_{k}(\bm{s})\exp\left[\phi(\bm{s})\right],
%\end{equation*}
One can then quickly apply a quadratic approximation to $L_w^{(k)}(f_k)$ to obtain
\begin{equation*}
\begin{split}
    L_w^{(k)}(f_k)
    &\simeq\Omega(f_k)-\sum_{\bm{x}\in \bm{\mathrm{x}}}\hat{w}_{k}(\bm{x})\hat{\phi}(\bm{x};\hat{\bm f}_{k-1})
    +\int_{\mathcal{S}}\hat{w}_{k}(\bm{s})\exp\left[\hat{\phi}(\bm{s};\hat{\bm f}_{k-1})\right]\mathrm{d}\bm{s}
    -\sum_{\bm{x}\in \bm{\mathrm{x}}}\hat{w}_{k}(\bm{x})f_k(\bm{x})\\
    &+\int_{\mathcal{S}}\hat{w}_{k}(\bm{s})\exp\left[\hat{\phi}(\bm{s};\hat{\bm f}_{k-1})\right]f_k(\bm{s})\mathrm{d}\bm{s}
    +\frac{1}{2}\int_{\mathcal{S}}\hat{w}_{k}(\bm{s})\exp\left[\hat{\phi}_{k-1}(\bm{s};\hat{\bm f}_{k-1})\right]f^2_k(\bm{s})\mathrm{d}\bm{s}.
\end{split}
\end{equation*}
Since $\hat{w}_{k}(\bm{s})$ and $\hat{\phi}(\bm{s};\hat{\bm f}_{k-1})$ are known, one can remove the constant terms to simplify it to
\begin{equation*}
\begin{split}
    \Tilde{L}_w^{(k)}(f_k)
    &\simeq\Omega(f_k)
    -\sum_{\bm{x}\in \bm{\mathrm{x}}}\hat{w}_{k}(\bm{x})f_k(\bm{x})
    +\int_{\mathcal{S}}\hat{w}_{k}(\bm{s})\exp\left[\hat{\phi}(\bm{s};\hat{\bm f}_{k-1})\right]f_k(\bm{s})\mathrm{d}\bm{s}\\
    &+\frac{1}{2}\int_{\mathcal{S}}\hat{w}_{k}(\bm{s})\exp\left[\hat{\phi}(\bm{s};\hat{\bm f}_{k-1})\right]f^2_k(\bm{s})\mathrm{d}\bm{s}.
\end{split}
\end{equation*}
Denote the set of locations where $\bm{z}(\bm{s})$ belonging to a leaf $v$ by $I_{kv}=\{\bm{s}:q(\bm{s})=v\}$. We can extract the contribution of this leaf to $\Tilde{L}_w^{(k)}(f_k)$, expand $\Omega(f_k)$ and replace $f_k$ by $\theta_{kv}$ to obtain
\begin{equation*}
\begin{split}
    \Tilde{L}_{wv}^{(k)}
    &=\gamma |\theta_{kv}|-\sum_{\bm{x}\in \bm{\mathrm{x}}}\hat{w}_{k}(\bm{x})1\{\bm{x}\in I_{kv}\}\theta_{kv}\\
    &+\int_{\mathcal{S}}\hat{w}_{k}(\bm{s})\exp\left[\hat{\phi}(\bm{s};\hat{\bm f}_{k-1})\right]1\{\bm{s}\in I_{kv}\}\left(\theta_{kv}+\frac{1}{2}\theta_{kv}^2\right)\mathrm{d}\bm{s}.
\end{split}
\end{equation*}
For any given tree structure $q_k$, it is straightforward to check that the minimizer of $\Tilde{L}_{wv}^{(k)}$, denoted by $\hat\theta_{kv}$, is
\begin{equation*}
\hat\theta_{kv} = \frac{\mathrm{sgn}(R_{kv}-T_{kv})\max\left(\left|R_{kv}-T_{kv}\right|- \gamma,0\right)}{T_{kv}},
\end{equation*}
where
\begin{equation*}
    R_{kv}=\sum_{\bm{x}\in \bm{\mathrm{x}}}\hat{w}_{k}(\bm{x})1\{\bm{x}\in I_{kv}\},\qquad T_{kv}=\int_{\mathcal{S}}\hat{w}_{k}(\bm{s})\exp\left[\hat{\phi}(\bm{s};\hat{\bm f}_{k-1})\right]1\{\bm{s}\in I_{kv}\}\mathrm{d}\bm{s}.
\end{equation*}

To approximate the integral throughout the derivation above, one may consider a numerical quadrature approximation. 

Suppose that the observation window $\mathcal{S}$ can be divided into $m$ grid cells $\mathcal{T}_1,\cdots, \mathcal{T}_m$, and each cell is centered at $\bm t_i$ and has a volume $|\mathcal{T}_i|$. Then, the following integral can be approximated as 
\begin{equation*}
    \int_{\mathcal{S}}\hat{w}_{k}(\bm{s})h(\bm{s})\mathrm{d}\bm{s}=\sum_{i=1}^m \hat{w}_{k}(\bm{t}_i)h(\bm{t}_i)|\mathcal{T}_i|,
\end{equation*}
where $h(\bm{s})$ is any measurable function defined on $\mathcal{S}$ such that $\hat{w}_{k}(\bm{s})h(\bm{s})$ is absolutely integrable.

\section{Supplementary Evaluation Results for the Simulation Study in Sec.~\ref{sec:simulation}}
\label{sec:Appendix_C}

In this section, we supply the results of the integrated absolute error for the main simulation study in Sec.~\ref{sec:simulation}. The integrated absolute error reads $\int_{\mathcal{S}}|\lambda(\bm{s})-\hat{\lambda}(\bm{s})|\mathrm{d}\bm{s}$, where $\lambda(\bm{s})$ and $\hat{\lambda}(\bm{s})$ denote the theoretical and estimated intensities, respectively. We report the results on Poisson process data in Tab.~\ref{tab:5} and those on log-Gaussian Cox and Neyman-Scott process data in Tab.~\ref{tab:6} and ~\ref{tab:7}.

In general, the findings convey similar messages to those reflected by Tab.~\ref{tab:1}, ~\ref{tab:2} and \ref{tab:3}. Specifically, in the scenario of Sec.~\ref{subsec:lowdimensionaldata} where there are a small number of covariates, XGBoostPP obtains comparable performance against existing approaches on all data types under various parameter settings. It outperforms KIEs while APPs behave the best. In the scenario of Sec.~\ref{subsec:highdimensionaldata} where there are a larger number of covariates, XGBoost significantly outperforms the APP method whose estimation error is about twice that of XGBoostPP. Comparing the estimation errors of XGBoostPP$_{\rm p}$ to XGBoostPP$_{\rm wp}$, the latter are substantially reduced on clustered processes while remain almost unchanged on Poisson processes. Moreover, such reductions are greater when point patterns are more spatially varying and clustered. 

In addition, comparing the scenario of Sec.~\ref{subsec:additionalsimulationstudy} to Sec.~\ref{subsec:lowdimensionaldata}, we find that, due to the presence of many nuisance covariates, the integrated absolute errors of both XGBoostPP and APP increase. However, the increase of XGBoostPP is much smaller than that of APP. This indicates that, while the true intensity function is simple, APP is much less robust than XGBoostPP when there is considerable uncertainty in choosing the most relevant covariates. 

%Last but not least, in both scenarios of Sec.~\ref{subsec:highdimensionaldata} and Sec.~\ref{subsec:additionalsimulationstudy}, the error reductions by XGBoostPP$_{\rm wp}$ against XGBoostPP$_{\rm p}$ when $\sigma=0.02$ are more prominent than when $\sigma=0.04$ for both log-Gaussian Cox and Neyman--Scott processes. %revealing that the estimation of clustering strength is a challenging task when complex covariate relationships exist. 

\begin{table*}[ht]
\caption{Averaged integrated absolute errors (standard deviations) of different approaches on Poisson process data.}
\label{tab:5}
\begin{center}
\begin{scriptsize}
\begin{sc}
\setlength{\tabcolsep}{4pt}
\begin{tabular}{p{0.8cm}p{2cm}ll}
\toprule
Covs & Poisson process & $\beta=0.5$ & $\beta=1.0$\\
\midrule
\multirow{6}{*}{\parbox{0.8cm} {\centering $2$\\ \centering (§4.1)}} & $\mathrm{KIE}_{\mathrm{ra}}$ & $91.1(10.2)$ & $170.0(8.9)$\\
& $\mathrm{KIE}_{\mathrm{re}}$ & $129.9(10.3)$ & $215.2(10.4)$\\
& $\mathrm{APP}_{\mathrm{naive}}$ & $\pmb{59.8}(11.0)$ & $\pmb{75.8}(11.1)$\\
& $\mathrm{APP}_{\mathrm{rfm}}$ & $60.7(10.6)$ & $77.7(10.6)$\\
& $\mathrm{XGBoostPP}_{\mathrm{p}}$ & $86.0(10.3)$ & $103.9(12.0)$\\
& $\mathrm{XGBoostPP}_{\mathrm{wp}}$& $86.1(10.3)$ & $103.7(12.2)$ \\
\cmidrule(lr){3-4}
\multirow{3}{*}{\parbox{0.8cm} {\centering $10$\\ \centering (§4.3)}} & $\mathrm{APP}_{\mathrm{rfm}}$ & $191.7(7.2)$ & $218.4(7.7)$\\
& $\mathrm{XGBoostPP}_{\mathrm{p}}$ & $\pmb{116.9}(12.0)$ & $\pmb{137.0}(11.3)$\\
& $\mathrm{XGBoostPP}_{\mathrm{wp}}$& $\pmb{116.9}(12.0)$ & $\pmb{137.0}(11.3)$\\
\midrule
Covs & Poisson process & $\beta=0.2$ & $\beta=0.4$\\
\midrule
\multirow{3}{*}{\parbox{0.8cm} {\centering $10$\\ \centering (§4.2)}}& $\mathrm{APP}_{\mathrm{rfm}}$ & $210.0(15.0)$ & $226.5(7.9)$\\
& $\mathrm{XGBoostPP}_{\mathrm{p}}$ & $\pmb{109.3}(10.2)$ & $149.8(11.9)$\\
& $\mathrm{XGBoostPP}_{\mathrm{wp}}$& $\pmb{109.3}(10.2)$ & $\pmb{149.8}(11.8)$\\
\bottomrule
\end{tabular}
\end{sc}
\end{scriptsize}
\end{center}
\end{table*}

\begin{table*}[ht]
\caption{Averaged integrated absolute errors (standard deviations) of different approaches on log-Gaussian Cox process data.}
\label{tab:6}
\begin{center}
\begin{scriptsize}
\begin{sc}
\setlength{\tabcolsep}{4pt}
\begin{tabular}{p{0.8cm}p{1.8cm}llll}
\toprule
\multirow{2}{*}{\parbox{0.8cm} {Covs}} & \multirow{2}{*}{\parbox{1.8cm} {Log-Gaussian\\ Cox process}} & \multicolumn{2}{c}{$\tau^2=1$} & \multicolumn{2}{c}{$\tau^2=2$} \\
& & $\sigma=0.02$ & $\sigma=0.04$ & $\sigma=0.02$ & $\sigma=0.04$\\
\midrule
& & \multicolumn{4}{c}{$\beta=0.5$} \\ \cmidrule(lr){3-6}
\multirow{6}{*}{\parbox{0.8cm} {\centering $2$\\ \centering (§4.1)}} & $\mathrm{KIE}_{\mathrm{ra}}$    & $93.7(14.7)$ & $99.5(19.6)$ & $96.8(18.0)$ & $110.8(29.8)$\\
& $\mathrm{KIE}_{\mathrm{re}}$    & $131.7(13.5)$ & $135.7(19.8)$ & $134.0(16.7)$ & $144.3(30.3)$\\
& $\mathrm{APP}_{\mathrm{naive}}$  & $\pmb{65.9}(15.4)$ & $78.0(23.0)$ & $74.9(20.0)$ & $101.7(38.2)$\\
& $\mathrm{APP}_{\mathrm{rfm}}$   & $66.3(15.0)$ & $\pmb{77.9}(22.9)$ & $\pmb{74.8}(19.7)$ & $\pmb{100.8}(37.7)$\\
& $\mathrm{XGBoostPP}_{\mathrm{p}}$ & $94.1(14.9)$ & $102.6(20.3)$ & $100.5(18.4)$ & $120.0(31.2)$\\
& $\mathrm{XGBoostPP}_{\mathrm{wp}}$ & $92.6(17.0)$ & $97.7(21.1)$ & $92.1(19.2)$ & $108.4(29.2)$\\
\cmidrule(lr){3-6}
\multirow{3}{*}{\parbox{0.8cm} {\centering $10$\\ \centering (§4.3)}} & $\mathrm{APP}_{\mathrm{rfm}}$   & $200.2(9.9)$ & $213.2(17.5)$ & $213.5(18.0)$ & $243.9(36.0)$\\
& $\mathrm{XGBoostPP}_{\mathrm{p}}$ & $118.8(16.1)$ & $126.9(19.8)$ & $125.8(20.3)$ & $155.0(36.6)$\\
& $\mathrm{XGBoostPP}_{\mathrm{wp}}$ & $\pmb{115.1}(16.8)$ & $\pmb{121.2}(19.8)$ & $\pmb{116.0}(17.0)$ & $\pmb{141.5}(26.2)$\\
\cmidrule(lr){3-6}
& & \multicolumn{4}{c}{$\beta=1.0$} \\ \cmidrule(lr){3-6}
\multirow{6}{*}{\parbox{0.8cm} {\centering $2$\\ \centering (§4.1)}} & $\mathrm{KIE}_{\mathrm{ra}}$    & $171.5(12.3)$ & $175.7(16.6)$ & $174.0(15.9)$ & $182.0(24.4)$\\
& $\mathrm{KIE}_{\mathrm{re}}$    & $216.6(13.9)$ & $220.7(21.6)$ & $219.2(18.6)$ & $226.9(34.8)$\\
& $\mathrm{APP}_{\mathrm{naive}}$  & $\pmb{84.0}(15.4)$ & $\pmb{98.7}(26.8)$ & $\pmb{97.2}(27.0)$ & $125.7(52.2)$\\
& $\mathrm{APP}_{\mathrm{rfm}}$   & $85.3(15.1)$ & $99.2(26.5)$ & $97.6(26.3)$ & $\pmb{124.7}(50.5)$\\
& $\mathrm{XGBoostPP}_{\mathrm{p}}$ & $113.3(17.3)$ & $127.1(25.4)$ & $123.8(23.1)$ & $146.2(40.7)$\\
& $\mathrm{XGBoostPP}_{\mathrm{wp}}$ & $108.6(18.2)$ & $121.8(26.8)$ & $110.0(18.9)$ & $131.8(33.6)$\\
\cmidrule(lr){3-6}
\multirow{3}{*}{\parbox{0.8cm} {\centering $10$\\ \centering (§4.3)}} & $\mathrm{APP}_{\mathrm{rfm}}$   & $232.0(12.6)$ & $247.4(23.0)$ & $250.8(25.5)$ & $280.1(50.7)$\\
& $\mathrm{XGBoostPP}_{\mathrm{p}}$ & $146.1(14.8)$ & $162.2(22.2)$ & $164.4(22.3)$ & $193.8(39.8)$\\
& $\mathrm{XGBoostPP}_{\mathrm{wp}}$ & $\pmb{142.2}(15.9)$ & $\pmb{160.2}(22.7)$ & $\pmb{147.5}(19.1)$ & $\pmb{179.1}(31.6)$\\
\midrule
& & \multicolumn{4}{c}{$\beta=0.2$}\\ \cmidrule(lr){3-6}
\multirow{3}{*}{\parbox{0.8cm} {\centering $10$\\ \centering (§4.2)}} & $\mathrm{APP}_{\mathrm{rfm}}$   & $208.0(12.2)$ & $214.2(14.2)$ & $214.2(14.9)$ & $240.9(32.6)$\\
& $\mathrm{XGBoostPP}_{\mathrm{p}}$ & $109.2(12.0)$ & $116.9(17.7)$ & $117.6(20.6)$ & $142.0(34.3)$\\
& $\mathrm{XGBoostPP}_{\mathrm{wp}}$ & $\pmb{104.9}(11.8)$ & $\pmb{112.9}(16.8)$ & $\pmb{110.5}(16.4)$ & $\pmb{135.6}(26.4)$\\
\cmidrule(lr){3-6}
& & \multicolumn{4}{c}{$\beta=0.4$}\\ \cmidrule(lr){3-6}
\multirow{3}{*}{\parbox{0.8cm} {\centering $10$\\ \centering (§4.2)}} & $240.9(32.6)$ & $234.4(10.2)$ & $246.1(17.5)$ & $246.5(17.0)$ & $272.7(40.1)$\\
& $\mathrm{XGBoostPP}_{\mathrm{p}}$ & $149.5(14.3)$ & $160.1(20.0)$ & $156.0(20.5)$ & $180.4(35.8)$\\
& $\mathrm{XGBoostPP}_{\mathrm{wp}}$ & $\pmb{140.9}(16.7)$ & $\pmb{154.3}(20.3)$ & $\pmb{139.7}(17.3)$ & $\pmb{166.9}(29.8)$\\
\bottomrule
\end{tabular}
\end{sc}
\end{scriptsize}
\end{center}
\end{table*}

\begin{table*}[ht]
\caption{Averaged integrated absolute errors (standard deviations) of different approaches on Neyman--Scott process data.}
\label{tab:7}
\begin{center}
\begin{scriptsize}
\begin{sc}
\setlength{\tabcolsep}{4pt}
\begin{tabular}{p{0.8cm}p{1.8cm}llll}
\toprule
\multirow{2}{*}{\parbox{0.8cm} {Covs}} & \multirow{2}{*}{\parbox{2cm} {Neyman--Scott\\ process}} & \multicolumn{2}{c}{$\kappa=100$} & \multicolumn{2}{c}{$\kappa=200$}\\
& & $\sigma=0.02$ & $\sigma=0.04$ & $\sigma=0.02$ & $\sigma=0.04$\\ 
\midrule
& & \multicolumn{4}{c}{$\beta=0.5$}\\ \cmidrule(lr){3-6}
\multirow{6}{*}{\parbox{0.8cm} {\centering $2$\\ \centering (§4.1)}} & $\mathrm{KIE}_{\mathrm{ra}}$    & $98.5(19.4)$ & $97.8(18.1)$ & $95.4(15.7)$ & $94.0(14.4)$\\
& $\mathrm{KIE}_{\mathrm{re}}$    & $136.8(19.0)$ & $135.6(18.8)$ & $132.7(15.0)$ & $132.0(14.4)$\\
& $\mathrm{APP}_{\mathrm{naive}}$   & $79.6(22.7)$ & $\pmb{72.8}(20.2)$ & $\pmb{69.2}(17.2)$ & $\pmb{66.3}(14.8)$\\
& $\mathrm{APP}_{\mathrm{rfm}}$   & $\pmb{79.3}(22.5)$ & $72.9(20.0)$ & $69.3(17.1)$ & $66.8(14.5)$\\
& $\mathrm{XGBoostPP}_{\mathrm{p}}$ & $103.9(19.3)$ & $99.9(19.0)$ & $96.8(16.1)$ & $94.1(14.6)$\\
& $\mathrm{XGBoostPP}_{\mathrm{wp}}$ & $95.3(20.1)$ & $95.4(20.5)$ & $92.4(18.4)$ & $93.6(15.5)$\\
\cmidrule(lr){3-6}
\multirow{3}{*}{\parbox{0.8cm} {\centering $10$\\ \centering (§4.3)}} & $\mathrm{APP}_{\mathrm{rfm}}$   & $223.3(16.0)$ & $209.6(13.5)$ & $206.8(11.3)$ & $199.5(9.8)$\\
& $\mathrm{XGBoostPP}_{\mathrm{p}}$ & $129.6(20.6)$ & $124.6(19.2)$ & $121.2(16.7)$ & $119.9(15.3)$\\
& $\mathrm{XGBoostPP}_{\mathrm{wp}}$ & $\pmb{119.5}(17.5)$ & $\pmb{119.9}(18.6)$ & $\pmb{114.7}(17.2)$ & $\pmb{119.4}(15.5)$\\
\cmidrule(lr){3-6}
& & \multicolumn{4}{c}{$\beta=1.0$}\\  \cmidrule(lr){3-6}
\multirow{6}{*}{\parbox{0.8cm} {\centering $2$\\ \centering (§4.1)}} & $\mathrm{KIE}_{\mathrm{ra}}$   & $175.3(17.4)$ & $174.5(14.6)$ & $173.6(13.3)$ & $172.3(11.7)$\\
& $\mathrm{KIE}_{\mathrm{re}}$    & $221.6(21.2)$ & $219.5(19.2)$ & $218.2(15.6)$ & $217.5(14.9)$\\
& $\mathrm{APP}_{\mathrm{naive}}$   & $104.9(27.0)$ & $\pmb{92.6}(23.2)$ & $\pmb{90.1}(19.3)$ & $\pmb{84.8}(16.4)$\\
& $\mathrm{APP}_{\mathrm{rfm}}$   & $\pmb{104.9}(26.8)$ & $93.6(22.7)$ & $91.2(18.8)$ & $85.9(16.1)$\\
& $\mathrm{XGBoostPP}_{\mathrm{p}}$ & $129.7(25.4)$ & $122.5(21.9)$ & $120.6(19.7)$ & $114.8(17.3)$\\
& $\mathrm{XGBoostPP}_{\mathrm{wp}}$ & $112.0(22.3)$ & $117.5(22.7)$ & $105.0(17.6)$ & $113.7(18.1)$\\
\cmidrule(lr){3-6}
\multirow{3}{*}{\parbox{0.8cm} {\centering $10$\\ \centering (§4.3)}} & $\mathrm{APP}_{\mathrm{rfm}}$  & $263.4(20.8)$ & $242.6(16.5)$ & $240.7(14.3)$ & $230.0(11.0)$\\
& $\mathrm{XGBoostPP}_{\mathrm{p}}$ & $172.2(22.3)$ & $155.9(20.1)$ & $153.1(17.1)$ & $146.1(15.8)$\\
& $\mathrm{XGBoostPP}_{\mathrm{wp}}$ & $\pmb{148.7}(22.2)$ & $\pmb{153.7}(20.1)$ & $\pmb{138.8}(17.9)$ & $\pmb{146.1}(15.8)$\\
\midrule
& & \multicolumn{4}{c}{$\beta=0.2$}\\ \cmidrule(lr){3-6}
\multirow{3}{*}{\parbox{0.8cm} {\centering $10$\\ \centering (§4.2)}} & $\mathrm{APP}_{\mathrm{rfm}}$   & $221.1(13.8)$ & $211.0(11.6)$ & $209.0(10.6)$ & $207.3(11.4)$\\
& $\mathrm{XGBoostPP}_{\mathrm{p}}$ & $120.0(19.4)$ & $114.8(16.2)$ & $110.0(13.3)$ & $110.9(13.4)$\\
& $\mathrm{XGBoostPP}_{\mathrm{wp}}$ & $\pmb{113.9}(17.0)$ & $\pmb{111.2}(15.3)$ & $\pmb{104.9}(13.1)$ & $\pmb{109.9}(13.1)$\\
& & \multicolumn{4}{c}{$\beta=0.4$}\\  \cmidrule(lr){3-6}
\multirow{3}{*}{\parbox{0.8cm} {\centering $10$\\ \centering (§4.2)}} & $\mathrm{APP}_{\mathrm{rfm}}$   & $253.9(14.7)$ & $241.1(12.0)$ & $238.7(11.1)$ & $232.7(9.2)$\\
& $\mathrm{XGBoostPP}_{\mathrm{p}}$ & $161.3(17.7)$ & $158.1(16.8)$ & $143.5(13.4)$ & $153.3(14.2)$\\
& $\mathrm{XGBoostPP}_{\mathrm{wp}}$ & $\pmb{143.6}(17.5)$ & $\pmb{153.0}(17.8)$ & $\pmb{138.4}(12.9)$ & $\pmb{153.2}(14.2)$\\
\bottomrule
\end{tabular}
\end{sc}
\end{scriptsize}
\end{center}
\end{table*}

\section{Flexibility of XGBoostPP Demonstrated by Three Poisson Toy Examples}
\label{sec:Appendix_D}

To intuitively visualize the flexibility of XGBoostPP in modeling various nonlinear relations between covariates and the point process intensity function, we demonstrate three Poisson toy examples. 

The examples are created on the unit window as $\mathcal{S}$ and take the $x,y$ coordinates as covariates. The first example has the intensity function $\lambda(x,y)=\exp(3+2x+7y)$, which is basically log-linear. The second example employs the intensity $\lambda(x,y)=\exp[8+16(x-0.5)^2]$, which aims to show the automatic variable selection capability of XGBoostPP. The third example considers the intensity function $\lambda(x,y)=\exp[2+4\mathrm{sin}(16x)+6\mathrm{sin}(16y)]$, which intends to show the capacity for modeling extraordinary covariate responses. 

\begin{figure}[ht]
\begin{center}
\centerline{\includegraphics[width=1.0\columnwidth]{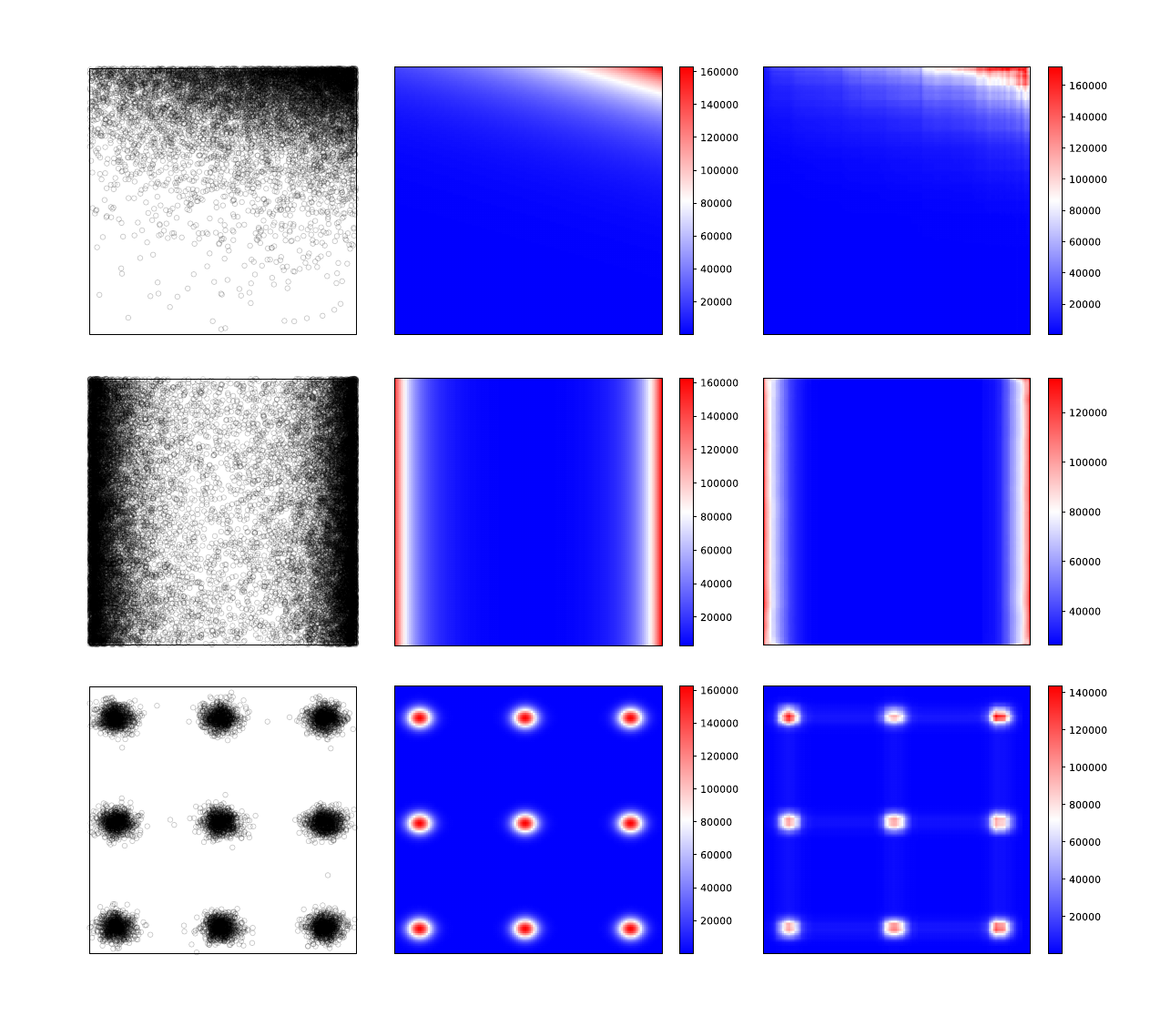}}
\vskip -0.2in
\caption{Point patterns, true intensities and the intensities estimated by XGBoostPP$_{\rm p}$ for the three toy examples.}
\label{fig:3}
\end{center}
\end{figure}

For all examples, we generate a point pattern based on the designed intensity function and fit XGBoostPP using the Poisson likelihood loss function to estimate the intensity as they are all Poisson processes. To test the flexibility, we only input $x,y$ as the spatial covariates to XGBoostPP and force it to learn the covariate relationships and the intensity trends on its own. We plot point patterns, true intensities and the estimated intensities by XGBoostPP in Fig.~\ref{fig:3}. The pictures clearly show that our XGBoostPP model can detect different covariate relationships and perform an automatic variable selection.

\section{Additional Details for the Real Data Analyses in Sec.~\ref{sec:realworldStudy}}
\label{sec:Appendix_E}

In this section, we supply additional experimental details for the real data analyses in Sec.~\ref{sec:realworldStudy}.

\subsection{Covariates Information in the Two Data Sets}
\label{sec:Appendix_E1}

The covariates considered in the tropical forest data set (`Bei', `Capp') and the kitchen fire data set (`Fire'), along with their descriptions, are listed in Tab.~\ref{tab:6} and \ref{tab:7}, respectively.

\begin{table*}[ht]
\caption{The covariates involved in the forestry data set (`Bei' and `Capp').}
\label{tab:8}
\vskip 0.1in
\begin{center}
\begin{small}
\begin{tabular}{lll}
\toprule
        Data & Covariate &  Description\\
         \midrule
         \multirow{8}{*}{\parbox{0.9cm} {Bei\\ Capp}} & Elev & The terrain elevation\\
         & Grad & The terrain slope\\
         & Cu & The content of Cu\\
         & Nmin & The content of Nmin\\
         & P & The content of P\\
         & pH & The pH value of soil\\
         & Solar & The solar index\\
         & Twi & The wetness index\\
\bottomrule
\end{tabular}
\end{small}
\end{center}
\end{table*}

\begin{table*}[ht]
\caption{The covariates involved in the fire data set (`Fire').}
\label{tab:9}
\vskip 0.1in
\begin{center}
\begin{small}
\begin{tabular}{lll}
\toprule
        Data & Covariate &  Description\\
         \midrule
         \multirow{29}{*}{\parbox{0.9cm} {Fire}} & House & The total number of houses\\
         & House\_indu & The number of houses with an industrial function\\
         & House\_hotl & The number of houses with a hotel function\\
         & House\_resi & The number of houses with a residential function\\
         & House\_$20$ & The number of houses constructed before $1920$\\
         & House\_$2045$ & The number of houses constructed in $[1920,1945)$\\
         & House\_$4570$ & The number of houses constructed in $[1945,1970)$\\
         & House\_$7080$ & The number of houses constructed in $[1970,1980)$\\
         &House\_$8090$ & The number of houses constructed in $[1980,1990)$\\
         &House\_$90$ & The number of houses constructed after $1990$\\
         &House\_frsd & The number of free standing houses\\
         &House\_other & The number of other houses\\
         &Resid & The number of residents\\
         &Resid\_$14$ & The number of residents with an age in $[0,14)$\\
         &Resid\_$1524$ & The number of residents with an age in $[15,24)$\\
         &Resid\_$2544$ & The number of residents with an age in $[25,44)$\\
         &Resid\_$4564$ & The number of residents with an age in $[45,64)$\\
         &Resid\_$65$ & The number of residents with an age over $65$\\
         &Man & The number of male residents\\
         &Woman & The number of female residents\\
         &Resid\_frsd & The number of residents living in free standing houses\\
         &Address & The density of addresses in the block\\
         &Urbanity & The urbanity of the block\\
         &Town & Boolean variable indicating the presence of a town\\
         &Poor & The percentage of poor residents (income $0-20$ percent)\\
         &Rich & The percentage of rich residents (income $80-100$ percent)\\
         &Value\_house & The average value of the houses in the block\\
         &Gas\_use & The average gas use in $m^3$ in the block\\
         &Elec\_use & The average electricity use in $kWh$ in the block\\
\bottomrule
\end{tabular}
\end{small}
\end{center}
\end{table*}

\subsection{The Test Poisson Log-likelihood Evaluation Metric}
\label{sec:Appendix_E2}

The evaluation metric used in the real data analyses is the cross-validated Poisson log-likelihoods of point patterns. Since we apply a four-fold cross-validation, the test Poisson log-likelihood over the data of one of the four subsets $\bm{\mathrm{x}}_i$, with $\bm{\mathrm{x}}=\{\bm{\mathrm{x}}_1,\dots,\bm{\mathrm{x}}_4\}$, reads
\begin{equation*}
    pl_{\mathrm{test},i}=\sum_{\bm{x}\in \bm{\mathrm{x}}_i }\log\left[\frac{1}{3}\hat{\lambda}_i(\bm{x})\right]-\frac{1}{3}\int_{\mathcal{S}}\hat{\lambda}_i(\bm{s})\mathrm{d}\bm{s},
\end{equation*}
where $\hat{\lambda}_i(\bm{s})$ is the estimated intensity function for fold $i$ based on the other three folds of training data. %Note that, here, $\hat{\lambda}_i(\bm{s})=3\hat{\lambda}(\bm{s})/4$ if we denote the estimated intensities fitted to the whole $\bm{\mathrm{x}}$ by $\hat{\lambda}(\bm{s})$, because we always use three quarters of the data of $\bm{\mathrm{x}}$ for training in each fold of the cros- validation. 
We sum up this test Poisson log-likelihood over all four subsets of $\bm{\mathrm{x}}_i$ to obtain the cross-validated test Poisson log-likelihood
\begin{equation*}
    pl_{\mathrm{test}}=\sum_{i=1}^4 pl_{\mathrm{test},i}=\sum_{i=1}^4 \left\{\sum_{\bm{x}\in \bm{\mathrm{x}}_i }\log\left[\frac{1}{3}\hat{\lambda}_i(\bm{x})\right]-\frac{1}{3}\int_{\mathcal{S}} \hat{\lambda}_i(\bm{s})\mathrm{d}\bm{s}\right\}.%= \sum_{i=1}^4 \sum_{\bm{x}\in \bm{\mathrm{x}}_i }\log\left[\frac{1}{4}\hat{\lambda}(\bm{x})\right]-\int_{\mathcal{S}} \hat{\lambda}(\bm{s})\mathrm{d}\bm{s}
\end{equation*}
%Based on that, the cross-validated test Poisson log-likelihood on the whole data of $\bm{\mathrm{x}}$ is computed by
%\begin{equation*}
%    pl_{\mathrm{test}}=\sum_{i=1}^4 pl_{\mathrm{test},i}+\sum_{\bm{x}\in \bm{\mathrm{x}}}\log(4).
%\end{equation*}

%%%%%%%%%%%%%%%%%%%%%%%%%%%%%%%%%%%%%%%%%%%%%%%%%%%%%%%%%%%%

\end{document}